\documentclass[onecolumn, amsmath, amssymb, superscriptaddress]{revtex4}

\usepackage{dcolumn}% Align table columns on decimal point
\usepackage{bm}% bold math
\usepackage{amsmath}
\usepackage{amsfonts}
\usepackage{mathrsfs}  
\usepackage{graphics}
\usepackage[dvips]{graphicx}
\usepackage{times}
\usepackage{setspace}
\usepackage{color}   % {\color{red} TEXT }
\usepackage{verbatim}
\usepackage[raggedright]{titlesec}
\usepackage[normalem]{ulem}
\usepackage{framed}
\usepackage{enumerate}
\usepackage[breaklinks=true, hyperfootnotes=false]{hyperref}
\usepackage{breakurl}

\usepackage[graphicx]{realboxes}
\usepackage{varwidth}
\usepackage{balance}
\usepackage{flushend}

\hypersetup{colorlinks=true,citecolor=blue, linkcolor=black,urlcolor=blue} % UNCOMMENT FOR SUBMISSION
\usepackage{hyperref}
\usepackage{amssymb}
\usepackage{mathbbol}
\usepackage{flushend}
\usepackage[outercaption]{sidecap} 

\newcommand{\beginsupplement}{%         
     \setcounter{page}{1} 
        \setcounter{section}{0} 
        \renewcommand{\thesection}{S\arabic{section}}%
        \setcounter{table}{0}
        \renewcommand{\thetable}{S\arabic{table}}%
        \setcounter{figure}{0}
        \renewcommand{\thefigure}{S\arabic{figure}}%
        \setcounter{equation}{0}
        \renewcommand{\theequation}{S\arabic{equation}}%
     }
     
\begin{document}
\title{The disruption index suffers from citation inflation \\ and is  confounded by shifts in scholarly citation practice} 
\author{Alexander Michael Petersen$^{a,}$}
\author{Felber Arroyave}
\affiliation{Department of Management of Complex Systems, Ernest and Julio Gallo Management Program, School of Engineering, University of California, Merced, California 95343, USA}
\author{Fabio Pammolli}
\affiliation{Politecnico Milano, Department of Management, Economics and Industrial Engineering, Via Lambruschini, 4/B, 20156, Milan, Italy}

\begin{abstract}
\noindent 
Measuring the rate of innovation in academia and industry is fundamental to monitoring the efficiency and competitiveness of the knowledge economy. To this end, a disruption index (CD) was recently developed and applied to publication and patent citation networks (Wu et al., {\it Nature} 2019; Park et al., {\it Nature} 2023).
%(Funk et al., {\it Management Science} 2017 \citep{funk2017dynamic}; Park et al., {\it Nature} 2023 \citep{park2023papers}).  
Here we show that CD systematically decreases over time due to secular growth in research and patent production, following two distinct mechanisms unrelated to innovation -- one behavioral and the other structural. Whereas the behavioral explanation reflects shifts associated with techno-social factors (e.g. self-citation practices), the structural explanation follows from `citation inflation' (CI), an inextricable feature of real citation networks attributable to increasing reference list lengths, which causes CD to systematically decrease. We demonstrate this causal link by way of mathematical deduction, computational simulation, multi-variate regression, and quasi-experimental comparison of the disruptiveness of PNAS versus PNAS Plus articles, which differ only in their lengths. Accordingly, we analyze CD data available in the SciSciNet database and find that disruptiveness incrementally increased from 2005-2015, and that the negative relationship between disruption and team-size is remarkably small in overall magnitude effect size, and shifts from negative to positive for team size $\geq$ 8 coauthors.
\end{abstract}

\maketitle

\centerline{{ \bf  Keywords:} Disruption index, innovation, secular growth, measurement bias, quasi-experiment} 

\bigskip

\noindent {\bf   One sentence summary: } The disruption index is  susceptible to measurement error,  which calls into question a growing body of research employing $CD$ to measure innovation trends and identify co-factors associated with team assembly. \\ 

\footnotetext[1]{ \ \ Send correspondence to:  apetersen3@ucmerced.edu}

\vspace{-0.3in}
\section*{Introduction}
\vspace{-0.2in}
Disruptive innovation refers to intellectual and industrial breakthroughs that sidestep conventional theory or practice by appealing to new value networks,  to the extent that the disruptive entrants  can quickly and unexpectedly  overcome the competitive advantages characteristic of established incumbents \cite{christensen2013disruptive}.
In the case of scientific advancement, the process of disruptive innovation manifests as intellectual contributions that appeal to novel  configurations of concepts and methods belonging to the  knowledge network \cite{pan2016memory,BMConvergence_2021,RecombInnov_2022},  
thereby  substituting prior combinatorial knowledge \`a la Schumpeter's theory of creative destruction \cite{schumpeter2013capitalism}.   
Against this backdrop,  scholars recently developed an index for quantifying citation disruption  (denoted by $CD$)      according to the implicit value of intellectual attribution that is encoded within the local  structure of citation networks  \cite{funk2017dynamic}, with the objective of identifying intellectual contributions that appeal to new streams of intellectual attribution while subverting established ones.  

Identifying the micro-level processes underlying disruption and quantifying its overall rate are fundamental to  understanding  scientific progress, and can in  principle  guide the management of institutions and  policies that accelerate innovation. As such, the $CD$ index  has received considerable  attention and inspired a significant volume of follow-up research. 
 However, there is a growing literature that challenges the  definition and application of $CD$ to real scientific and patent citation networks \cite{bornmann2020disruption,ruan2021rethinking,petersen2023disruption,macher2023illusive,bentley2023disruption,leibel2023we,holst2024dataset}.  
One stream  of critique calls into question the long-term temporal decline in $CD$ reported by \cite{park2023papers}, and the morose interpretation of its implications on the status and outlook of the scientific endeavor \cite{kozlov2023disruptive,holst2024dataset}. 
In particular, Macher et al.  \cite{macher2023illusive} identify a substantial number of missing patent citations in the data analyzed in   \cite{park2023papers}, both at the beginning of their data sample and towards the end, which effectively reduces the number of backwards (i.e. references) and forward citations that were analyzed.  
 Once correcting for the data omission, the re-analysis reveals an {\it increasing rate} of disruption  in several patent domains central to the techno-informatic revolution of the last 30 years \cite{BMConvergence_2021}. 
 
 Similarly, a recent and independent re-analysis by Holst et al. \cite{holst2024dataset} shows that missing citations in the scientific publication data, which are more prominent in the early years of the sample, give rise to a subsample of publications with 0 references (which by definition correspond to maximum disruption value of $CD =1$). They show that the prevalence of these  temporally biased data anomalies are entirely sufficient to generate the negative trend in $CD$ reported by  \cite{park2023papers}; upon correcting for these anomalies, they show that $CD$ trends  insubstantial for both patents and publications.     
 A  third independent study of scientific publications by Bentley et al. also report an accelerating rate of disruption at the end of their data sample after developing a weighted variant of $CD$ that accounts for  temporal shifts in the connectivity of real citation networks  \cite{bentley2023disruption}.  In addition to these studies  reporting an increasing rate of innovation according to temporal patterns in the citation network, a complementary approach based upon measuring combinatorial innovation in the knowledge network also reports a persistent innovation rate over time  \citep{RecombInnov_2022}.

A second stream of critique focuses on how  $CD$ is defined,  and the implications of its mathematical formulation on its reliability in  bibliometric analysis \cite{wu2019confusing,bornmann2020disruption,leydesdorff2021proposal,ruan2021rethinking,petersen2023disruption}. 
Taken together, these considerations  further call into question a number of studies reporting statistical relationships between $CD$ and various covariates  related to research production and team assembly   -- such as career productivity, team size, citation impact,  and the geographic dispersion of team members  \cite{li2024productive,wu2019large,park2023papers,wang2023evaluating,Lin_remote_2023,li2024productive} -- most notably because  each of these covariates also grows with time. Statistical relationships between $CD$ and other time-dependent covariates are susceptible to  omitted variable bias, which is a formidable source of measurement error in statistical analysis. As such, the connection between these  two streams is the role of secular growth,  which manifests as a temporal bias that underlies the data artifacts generating declining trends in $CD$ and the susceptibility of  $CD$ to measurement error due to its non-linear dependence on the structure and rate of backwards citations.

 Against this backdrop, here we contribute to these two streams by demonstrating how systematic measurement bias deriving from the inextricable densification of empirical citation networks, combined with omitted variables capturing confounding shifts in scholarly citation practice, further contributes to the mismeasurement of a  ``decline in disruptiveness'' \cite{park2023papers}. Our critique is centered upon the role of reference list  in the definition of $CD$, and the implications of  multifold increases in reference list lengths over the last half century,  which is a fundamental source of `citation inflation' (CI) \cite{petersen_citationinflation_2018}. Specifically, we apply four complementary methodologies that expose the underlying bias in  $CD$ definition and its application -- deductive quantitative reasoning, computational modeling, a quasi-experimental test, and multivariate regression -- the results of which are consistent with a companion study \cite{petersen2023disruption} based upon different data sources and regression model specification. 
 
In order to  test and validate the CI hypothesis  --   that increasing reference list lengths confound the measurement of trends in $CD$  and covariate relationships --  we developed a quasi-experiment based upon the entire corpus of  research published in  {\it Proceedings of the National Academy of Sciences of the United States of America} (PNAS) over the 5-year period 2011-2015. Our identification strategy is based around  comparing  research articles published in the traditional format (print and online publication) to those published as long-form  {\it PNAS Plus} articles (online-only publication) \cite{verma2012pnas,schekman2010creating} -- as these two publication formats are nearly indistinguishable, aside from the longer reference list lengths of {\it PNAS Plus} articles. 
We conclude with a large-scale multivariable regression analysis of 7.8 million articles from 1995-2015, which accounts for the data quality and measurement biases identified, thereby improving upon the methodological designs employed in \cite{wu2019large,park2023papers}. 
Results show that  (a) the net effect size  of temporal and team-size trends are at the level of noise and therefore inconsequential to science and innovation policy; and (b) by including appropriate controls and focusing on the metric itself instead of percentile values, the relationship between $CD$ and team size is instead increasing.

%\bigskip
%\centerline{[Figure 1 Here]}
%\bigskip

\vspace{-0.2in}
\section*{Background \& Related Literature \label{sec:liter_hyp}}
\begin{figure*}
\centering{\includegraphics[width=0.79\textwidth]{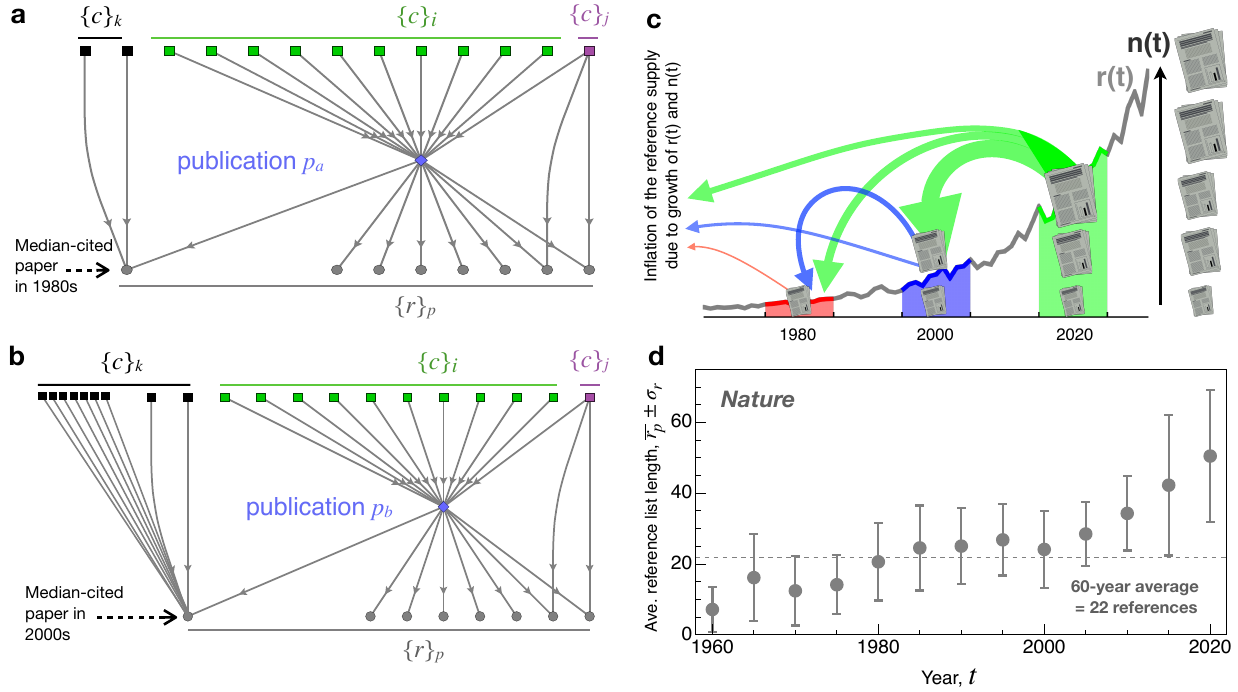}}
 \caption{  \label{Figure1.fig} {\bf  Citation inflation is an inextricable feature of citation networks.} 
 The disruption index $CD_{p}$ is calculated according to three non-overlapping subsets of $\{c\}_{p} = \{c\}_{i} \cup \{c\}_{j} \cup \{c\}_{k}$, of sizes $N_{i}$, $N_{j}$ and $N_{k}$, respectively. 
{\bf (a,b)} Schematic of the citation network sub-graph contributing to the calculation of the disruption index   for two papers that differ only in the connectivity of the single reference contributing to $N_{k}$. Moreover, in order to convey the magnitude and impact of secular growth as it manifests on real citation networks, 
the subset $\{c\}_{k}$ for publication $p_{a}$ is characteristic of  citation rates in the 1980s, whereas for  $p_{b}$ it is characteristic of the 2000s. Consequently, $CD_{a} = 0.69$ and $CD_{b} = 0.45$,  corresponding to a 35\% decrease in $CD$ attributable to 20 years of increasing citation network density. 
{\bf (c)} Schematic illustrating the inflation of the reference supply  owing to the fact that the annual publication rate $n(t)$ (comprised of increasingly variable article lengths), along with the number of references per publication $r(t)$, have grown exponentially over time. 
Consequently,  the observed densification is both within- and across-generation, such that older publications can receive more citations from present day research than from contemporaneous research due to secular growth. 
{\bf (d)} Citation inflation even affects journal with relatively small change in $n(t)$, such as traditional print journals like {\it Nature}, which have witnessed 7-fold increases in reference list lengths over the last 60 years.  }
\end{figure*}

\vspace{-0.2in}
\noindent {\bf Definition of CD and its susceptibility to secular growth}\\
\noindent  The disruption index $CD_{p}$  \cite{funk2017dynamic,park2023papers} measures the degree to which an intellectual contribution $p$ (e.g. a patent or academic research publication)  supersedes the sources cited in its reference list, denoted  by  the set $\{r\}_{p}$. 
The argument for $CD_{p}$ is that if future  contributions cite $p$ but do not cite members of $\{r\}_{p}$, then $p$ plays a disruptive role in the citation network. 
As such,  disruption can be inferred according to the local  structure of the subnetwork $\{r\}_{p}\cup p \cup  \{c\}_{p}$ that includes  the set of citing nodes  $\{c\}_{p}$  connecting to either the focal node $p$ or any member of $\{r\}_{p}$. 
According to its  definition   \cite{funk2017dynamic,park2023papers} reformulated   as a ratio  \cite{wu2019large},  $CD_{p}$ is calculated by identifying three non-overlapping subsets of $\{c\}_{p} = \{c\}_{i} \cup \{c\}_{j} \cup \{c\}_{k} $, of sizes $N_{i}$, $N_{j}$ and $N_{k}$, respectively -- see {\bf Fig. \ref{Figure1.fig}}(a,b)  for a schematic illustration. In practice, a citation window (CW) is used to temper the effects of right-censoring bias, such that only citations occurring within a CW-year period are included in the subnetwork $\{c\}_{p}$. In what follows, we employ a CW = 5-year window  denoted by $CD_{p,5}$, as in prior research \cite{park2023papers,wu2019large}; however, the fundamental issues with the definition of $CD$ are independent of  CW \cite{petersen2023disruption}, and so for brevity we represent the general definition by $CD_{p}$.  

The  subset $i$ refers to members of $\{c\}_{p}$ that cite the focal $p$ but do not cite any elements of $\{r\}_{p}$, and thus measures the degree to which $p$  disrupts the flow of attribution to  members of $\{r\}_{p}$. The  subset $j$ refers to  members of $\{c\}_{p}$ that cite both $p$ and $\{r\}_{p}$, measuring the degree of consolidation that manifests as triadic closure in the subnetwork (i.e.,  triangles  formed between $\{r\}_{p}$, $p$,  $\{c\}_{j}$). The subset $k$ refers to members of $\{c\}_{p}$ that cite $\{r\}_{p}$ but do not cite $p$. 
As such,  \cite{wu2019large}  show that  $CD_{p}$ can be calculated as the ratio
\begin{eqnarray}
\label{eqnCD}
CD_{p} = \frac{N_{i}-N_{j}}{N_{i}+N_{j}+N_{k}}   = \frac{CD^{\text{nok}}}{1+R_{k}} \ , % =  \frac{(N_{i}-N_{j})/(N_{i}+N_{j})}{1+N_{k}/(N_{i}+N_{j})}
\end{eqnarray}
where  the  second equivalence is a simple re-organization  of the equation to highlight the extensive quantity $R_{k} = N_{k}/(N_{i}+N_{j}) \in [0, \infty)$, which measures the rate of extraneous citation. 
Such re-organization  facilitates deducing the scaling behavior of $CD$ associated with the number of articles $n(t)$ and the average reference list length per article $r(t)$ per year $t$.
According to the scaling of network growth,    $R_{k} \propto r(t)$ because $N_{k} \sim n(t)r(t)$ and  $N_{i}+N_{j} \sim n(t)$, which is empirically validated in  \cite{petersen2023disruption}.
Hence, because the ratio $CD^{\text{nok}} = (N_{i}-N_{j})/(N_{i}+N_{j}) \in [-1,1]$ is an intensive  measure, $CD_{p}$ thus features a numerator that is bounded and a denominator that is unbounded -- and so $CD_{p}(t)$  converges to 0 over time  because  $R_{k}$ grows proportional $r(t)$.
Moreover, even alternate disruption definitions such as $CD^{\text{nok}}$ are biased, because shifts in scholarly practice that manifest as network autocorrelations, such as self citation and journal impact-factor boosting, increase the overall rate of consolidation (triadic closure) measured by the term $N_{j}$. \\

\bigskip
\noindent{\bf Citation inflation: a measurement bias deriving from secular growth}\\
\noindent Citation inflation (CI) refers to the exponential growth of citations produced via the secular growth of the scientific endeavor \cite{petersen_citationinflation_2018}.  
 {\bf Figure \ref{Figure1.fig}(c)} illustrates how CI  arises through the combination of increasing reference lists, denoted by $r(t)$, and increasing publication (or patenting) rates, denoted by $n(t)$, which significantly increases the density of citation networks over time. 
In the case of scientific research production, empirical growth rates estimated from the entire Clarivate Analytics Web of Science citation network  show that  total volume  of  citations  generated by the scientific literature, $C(t)$, grows exponentially with annual rate   $g_{C} = g_{n} + g_{r}  = 0.051$; %0.033 + 0.018
hence,  the number of links in the citation network is growing by roughly 5\% annually, corresponding to a doubling period of just  $\ln(2)/g_{C} = 13.6$ years \cite{pan2016memory}.

Moreover, as references tend to increasingly extend further back in time \cite{pan2016memory}, the impacts of CI are not constrained to contemporaneous layers of the citation network, but instead are cross-generational.
As such, the increasing density of  citation networks manifests at both the source (reference) and destination (citation) of each new link, which is a temporal bias that is challenging to  neutralize in the development of standardized network metrics.  
For example, in the 1980s the median-cited paper received 2 citations within 5 years of publication; however by the 2000s, this nominal quantity increased to 9 \cite{pan2016memory}. 
In addition,  there has been a paradigm shift towards online publishing that has facilitated  greater publication volumes, faster publication times,  and longer articles with longer reference lists. 
Take for example the journal {Nature} for which $n(t)$ has been roughly constant over the last 60 years: in 1970 the average number of references per articles was $\overline{r_{p}}=7$; by 2000 $\overline{r_{p}}$ increased to 24; and by 2020 $\overline{r_{p}} = 51$, a 7-fold increase over the 60-year period  -- see {\bf Fig. \ref{Figure1.fig}}(d) and {\bf Fig. \ref{FigureS1.fig}} in the {\it Supplemental Information}.\\ %, which shows that the persistent trend is not attributable to outliers, but rather a systematic shift towards larger $r_{p}$ values. 

\noindent{\bf Issues with prior works analyzing trends in CD}\\
Park et al.  \cite{park2023papers} develop four   robustness check approaches: comparison with alternative definitions of $CD$, normalization of $N_{k}$ in $CD$, regression adjustment, and synthetic randomization of the citation network. Yet each is susceptible to either data quality issues that are temporally biased towards early years, or measurement bias and omitted variable bias associated with the definition of $CD$. 

First,  because alternative disruption variants, namely  $CD^\text{nok}$ and $CD^{*}$ \cite{bornmann2020disruption,leydesdorff2021proposal},  are constructed around similar ratios, they are also susceptible to data quality issues as well as CI. 
However, it is notable that the alternative indices are less susceptible to CI, and indeed their trends shown in Extended Data Fig. 7 of  \cite{park2023papers}  are markedly less prominent than what is shown for $CD$. In the particular case of $CD^\text{nok}$, the location of the average value is large enough that a vast majority of papers are classified as disruptive -- which begs for improvement. Moreover, in our companion study focusing upon a computational model, we show that even the tempered decline in  $CD^\text{nok}$ can be attributed to CI \cite{petersen2023disruption} . 

Second,  in order to attenuate the effect of CI, Park et al.  \cite{park2023papers}  develop both  `paper' and `field x year' normalized variants of $CD$ by modifying the factor $N_{k}$ in Eq. (\ref{eqnCD}). In the first case, they replace $N_{k}$ with $N_{k}-r_{p}$. However, according to  scaling behavior arguments,  since $N_{k}\sim n(t)r(t)$ then  $(N_{k}-r_{p}) \sim (n(t)r(t)-r_{p}) \approx (n(t) -1)r(t)$, which does not generate the intended consequence.  Moreover, the reduction of $N_{k}$ by $r_{p}$ still renders these normalized variants  susceptible to the  scenario exhibited in {\bf Fig. \ref{Figure1.fig}}(a,b), whereby citing just a single highly-cited paper causes $N_{k}$ to vastly exceed the difference $N_{i}-N_{j}$ in the numerator of $CD$, such that $CD_{p}\rightarrow 0$.  In the second case of the field-year normalized variant, the average $r(t)$ for papers from the same field and year are subtracted, which  amounts to the same issue, $N_{k}-r(t)\sim n(t)r(t)-r(t) = (n(t) -1)r(t)$.

Third, the regression adjustment implemented by \cite{park2023papers} is poorly documented, as there is no model specification; moreover,   Extended Table 8 only shows the estimated coefficients for the year indicator variable, and  does not show the estimates for other controls. And according to Extended Data Table 1 in \cite{park2023papers}, their model specification does not incorporate  available publication-level factors that co-vary with  $CD_{p}$, namely $r_{p}$, $c_{p}$ and the number of coauthors, $k_{p}$.

And finally, robustness checks based upon rewired   citation networks are insufficient since the degree-preserving randomization holds constant $r_{p}$ or $c_{p}$. Hence, this randomization scheme can only be expected to attenuate biases  attributable to correlated citation behavior -- contrariwise, biases deriving from data quality issues and CI can be expected to persist.
Moreover, because  shuffling the citation network reduces the rate of triadic closure to random chance, then this null model essentially converts all $N_{j}$ links into $N_{i}$ links. Consequently, the expected randomized value is $CD^\text{Rand}_{p} = ((N_{i}+N_{j})-0)/(N_{i}+N_{j}+N_{k}) =  1 / (1+R_{k}) >0$, which is positive definite and converges to 0 as $R_{k}$ increases.  Park et al. then calculate a Z-score comparing the real and randomized values, $Z_{p}=(CD_{p} - CD^\text{Rand}_{p})/\sigma[{CD^\text{Rand}_{p}}]$, and plot the average value over time in  Extended Data Fig. 8  \cite{park2023papers}. Their results show extremely negative $Z$-score values (upwards of $2\sigma$ effect sizes). These deviations are  also  methodological artifacts: since $CD^\text{Rand}_{p}>0$, then all papers with $CD_{p}<0$ deterministically yield $Z_{p}<0$; moreover,  the standard deviation $\sigma[CD^\text{Rand}_{p}]$ is  extremely small because the chances of the randomization producing triadic closure is extremely small. Hence,  with little variation to work with around a systematically small value  $CD^\text{Rand}_{p} \approx 1 / (1+R_{k}) \sim 1/ r(t)$ ,  this randomization approach vastly underestimates the intrinsic scale of variation, i.e., $\sigma[CD^\text{Rand}_{p}] \ll \sigma[CD_{p}]$.  

In a different study on the relationship between $CD$ and team size,  Wu et al.  \cite{wu2019large} also develop robustness checks  based upon multi-variate regression. However there is no clear model specification provided in their Supplementary Table 4; hence,  in addition to omitting $c_{p}$ and $r_{p}$, it is unclear how they controlled for publication year. Moreover, the majority of their analysis is based upon descriptive trend analysis using {\it percentile} values of $CD$.  This mapping of nominal $CD$ values to percentiles  obfuscates the extremely narrow distribution of $CD$. At the same time, this modification of dependent variable generates the appearance of considerable effect sizes. Because most publications are concentrated around relatively small $CD$ values, a  small idiosyncratic shift in $CD$ will generate disproportionately large shifts in the percentile value. 

A third study analyzes the    relationship between $CD$ and collaboration distance among coauthors  \cite{Lin_remote_2023}. This analysis also omits $c_{p}$ and $r_{p}$ from their multivariate robustness check (see Extended Data Table 1), and instead categorize papers as being before or after 2000 (a crude temporal control) and being solo-author or not (a crude team size control). 
As an example of persistent negligence for confounding factors, Lin et al. write: ``For example, the 1953 paper on DNA by Watson and Crick is among the most disruptive works (D = 0.96, top 1\%), whereas the 2001 paper on the human genome by the International Human Genome Sequencing Consortium is highly developing (D = -0.017, bottom 6\%).'' Yet these papers are from vastly different socio-technological eras, with the former produced by two coauthors citing $r_{p} = 6$ prior works,  where  the latter  is attributed to 200+ coauthors and cites $r_{p} = 452$ prior works. 

Additionally,  results reported in \cite{Lin_remote_2023} are based upon  the relative rates of $CD>0$ versus $CD<0$. This dependent variable is simply the sign of $CD$,  which only depends on the numerator difference, $N_{i}-N_{j}$. While this choice may at first appear to be less susceptible to CI, it is still susceptible to  the data quality issues  identified by \cite{macher2023illusive,holst2024dataset}, as well as confounding trends in the rate of triadic closure attributable to shifts in self-citation and other correlated citation behavior, which are more prominent in larger teams -- and  larger teams are  more likely to be extended across larger distances.  

% data quality \& methodology
In response to these issues, two  streams of critique have emerged regarding research on $CD$ -- one regards data quality issues and the other regards methodological issues.
Regarding the former, citation networks based upon publication and patent data are susceptible to missing references, citations, and the classification of non-research oriented content (e.g. editorials) as the products of dedicated research. These data quality issues are more frequent for older publications, and less so for newer ones, as the modern publication industry benefits from information system features that were not available in the past (e.g. Digital Object Identifiers, and the web-based publication).
As demonstrated by \cite{holst2024dataset}, the frequency of publications and patents with 0 references is highly concentrated during early years. 
They show that this systematic bias in data quality contributes significantly to the decline in $CD$, since papers and patents with $r_{p}=0$ correspond to maximum disruption, $CD_{p} = (N_{i} - 0)/(N_{i}+0+0) =1$. Similarly, \cite{macher2023illusive} show that left-censoring bias in US patent data means that patents from early years are artificially missing  references to  patents before the starting date of the dataset; upon correcting for these omitted references, which increases $r_{p}$ closer to their true value, then the decline in $CD$ for patents is greatly reduced. 

% Methodology
A second stream of critique regards the methodological choices, e.g. omitted variables and the susceptibility of $CD$ to secular growth.
By way of example, Bentley et al.  \cite{bentley2023disruption} modify the definition of $CD_{p}$ to account for CI according to both the number of publications $n(t)$ and the total number of citations produced, $C(t) = n(t)r(t)$. Their re-analysis reveals an increasing weighted $CD$ from the early 1990s through 2013. They also critique the results of \cite{park2023papers} based upon natural language analysis, which also is susceptible to secular  trends affecting the usage frequency and fashion of  words. 

To summarize, several recent analyses  report findings of the  following form: as $X$  increases, $CD$ decreases (where $X$ is time,  individual publication rate, team size, nominal citations, and collaboration distance)  \cite{li2024productive,wu2019large,park2023papers,wang2023evaluating,Lin_remote_2023}. 
Accordingly, we conjecture that any  variable $X(t)$ that increases over time will generate correlations of this  pattern -- yet the degree to which such correlations survive  confounders and wether they represent significant effect sizes is a more intriguing matter.
To this end, here  we seek to consolidate  a growing number of critiques -- first by addressing methodological issues, and concluding with a regression framework that also addresses the data quality issues. 

\vspace{-0.2in}
\section*{Results \label{sec:results}}
\vspace{-0.2in}
\noindent {\bf Computational model incorporating CI}\\
\noindent We use a tested and validated  mechanistic citation network growth  model \cite{pan2016memory} to compare average trends in $CD$ calculated for synthetic networks   generated with and without  CI, which  are otherwise identical in their construction. This generative citation network model  belongs to the class of growth and redirection models \cite{krapivsky_network_2005,barabasi2016network} and  implements stochastic link dynamics that mimic preferential attachment \cite{barabasi2016network}, while also incorporating  other  latent features of scientific production,  in particular the exponential growth of $n(t)$ and $r(t)$.  This  model reproduces a number of statistical regularities established for real citation networks -- both structural (e.g. a log-normal citation distribution \cite{UnivCite}) and dynamical (e.g., increasing reference age with time \cite{pan2016memory}; exponential citation life-cycle decay \cite{petersen_reputation_2014}). 

 Network growth in this model is governed by  two complementary citation mechanisms that can be completely controlled by tunable parameters: (i) direct citation and (ii) redirected citation  \cite{pan2016memory,petersen2023disruption}. The second mechanism (ii) gives controls the rate of triadic closure in the synthetic citation network, thereby capturing correlated shifts in  scholarly citation practice, such as   citation trails illuminated by web-based hyperlinks that make it easier to find and cite prior literature; and self-citation among individuals and journals aimed at increasing their prominence in the attention economy. Moreover, the   `consolidation' measured by $N_{j}$ in $CD_{p}$ explicitly measures the number of citations that feature triadic closure. 
 In  related work focusing on the details of this computation model   \cite{petersen2023disruption}, we find that    CI has a stronger role than  redirection  in explaining the  decline in $CD_{5}(t)$, and so in what follows we focus on the effects of CI.
 
Accordingly, here we focus on the network growth parameters that determine the rate of CI, which thereby facilitates measuring the impact of secular growth  on two  trends: (a) the decline in  the average $CD_{p,5}$ value in year $t$, denoted by   $CD_{5}(t)$; and (b) the scaling hypothesis connecting the rate of extraneous citations featured in the denominator of Eq. (\ref{eqnCD})  to the average number of references per paper, $R_{k}(t) \propto r(t)$. Note that the increase in $r(t)$ manifests from secular growth  as well as shifts in scholarly citation practice. For example,  papers with more authors tend have longer reference lists  \cite{petersen2023disruption},  partly because larger teams tend to write longer papers, but also because there are more authors seeking to benefit from self-citation.

Hence, we use the computational model to the CI hypothesis by generating two distinct network ensembles, each comprised  of 10 random networks  grown over $t=1...T $  periods (representative of publication years), terminating the network growth at $T \equiv 150$. Each network realization is seeded with common initial conditions -- e.g. the initial cohort features $n(1)=30$ nodes, each with $r(1)=5$ references. These synthetic citation networks are available for cross-validation, and can be used to develop alternative scientometrics that are neutral to CI \cite{DryadDisruption2023}.

\begin{figure*}
\centering{\includegraphics[width=0.99\textwidth]{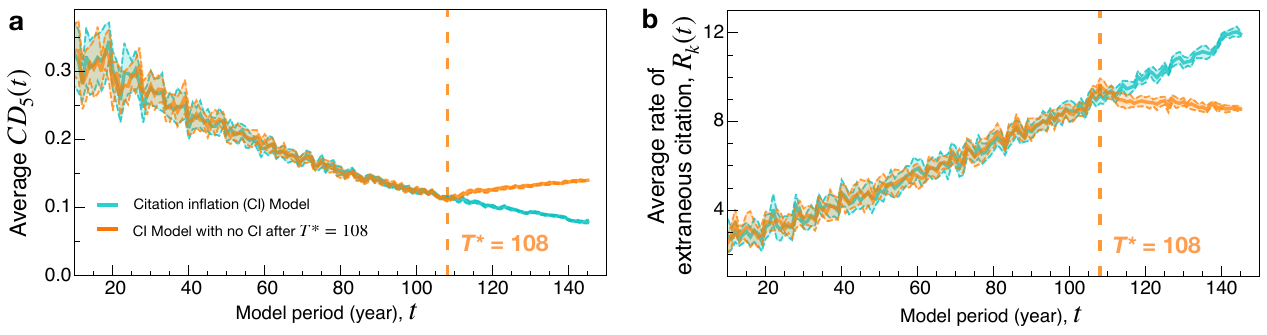}}
 \caption{  \label{Figure2.fig} {\bf  Computational simulation of growing citation networks: after `turning off' CI, the systematic decline in $CD$ reverses.} 
{\bf (a)} The average $CD_{5}(t)$ calculated across 10 different computational realizations of (i) the standard CI model and (ii) the CI model with quenched reference list growth ($g_{r}=0$) for $t\geq 108$. 
{\bf (b)} Average rate of extraneous citation, $R_{k}(t)$, showing that $CD_{5}(t)$ converges to 0 because the denominator of the  disruption index in  Eq. (\ref{eqnCD}) is unbounded as $r(t)$  grows. 
}
\end{figure*}

By construction, both network ensembles are statistically identical for the first 107 periods -- see {\bf Fig. \ref{Figure2.fig}}. The first ensemble incorporates the empirical  rates $g_{n}=0.033$ and  $g_{r}  = 0.018$ for the entire $T \equiv 150$ periods, with   networks reaching a final size of $N = 125,270$ nodes  and 5,948,492 links with  $r_{p} \equiv r(t=150) = 73$ after 150 periods. The second ensemble features the same  $g_{n}$ and thus reaches the same final size of $N = 125,270$ nodes. However, the growth of $r(t)$ is suddenly quenched at $T^{*}=108$ to  $g_{r}=0$, such that $r(t) = 34$ for $t\geq T^{*}$, which effectively `turns off' CI attributable to growing reference lists  -- see  {\bf Fig. \ref{FigureS2.fig}(a,b)}. This sudden hault represents a hypothetical scenario in which journals were to impose a hard cap on reference list lengths in order to temper  CI. Such caps are not inconceivable, as {\it Nature} provides a soft policy that ``articles typically have no more than 50 references'' in their \href{https://www.nature.com/nature/for-authors/formatting-guide}{formatting guide} for authors. 

{\bf Figure  \ref{Figure2.fig}}(a) compares the average $CD_{5}(t)$  for these two scenarios, which reproduces the magnitude of empirical decline reported in \cite{park2023papers} for $t<T^{*}$. However, the two curves diverge for $t\geq T^{*}$, with the curve featuring  quenched reference lists suddenly reversing course, thereby revealing the acute effect of CI on $CD$. Similarly, {\bf Fig.  \ref{Figure1.fig}}(b) tracks the growth of $R_{k}(t)$, showing that this quantity is extensive when $r(t)$ is growing, and intensive when $r(t)$ is constant. 
{\bf Figure  \ref{FigureS2.fig}}(c-f) confirm the relationship $R_{k}(t) \propto r(t)$, and show that the proportionality is  independent of the citation window (CW) used for calculating $CD_{CW}(t)$;  see \cite{petersen2023disruption} for additional empirical validation based upon a different dataset. Accordingly, our simplified network model demonstrates that CI fully controls the trends in $CD(t)$. \\

\noindent{\bf Quasi-experimental validation of the citation inflation hypothesis: Empirical analysis of $\vert CD_{p,5} \vert$} \\
\noindent Computational `toy models' are designed to capture the essential parameters underlying observed variation, while neglecting those features that are perceived to be  non-essential. 
However, an unavoidable limitation to such approaches is determining what exactly are the essential parameters. 
For example, in our parsimonious growth model we do not account for various sources of heterogeneity underlying scientific publication, such as team size and reference list lengths, which both extending from just a few  to several hundreds. 
Instead, all synthetic publications in our model from the same year have the same number of references, $r_{p} = r(t) \equiv $ average reference list length, so that we can rule out variation in this feature as a cause for the effect we are seeking to understand.   Hence, in order to further test the CI hypothesis in an empirical setting, we again resort to a simplified scenario.

Specifically, we  exploit the 2011 launch of a strategic publishing model developed by the journal  PNAS, consisting of a long-form online-only publishing option -- called {\it PNAS Plus} -- to complement its traditional print  option  \cite{schekman2010creating}.
Article submissions were not processed, reviewed or prioritized according to the author-designated print option \cite{verma2012pnas}, and so publications in these two formats are satisfactory counterfactuals for testing wether publications with larger $r_{p}$ are biased towards smaller $CD_{p}$. 
To demonstrate this causal link, we juxtapose the two sets of articles that are otherwise indistinguishable, on average, aside from  {\it PNAS Plus} articles having longer reference lists. 

\begin{figure*}
\centering{\includegraphics[width=0.99\textwidth]{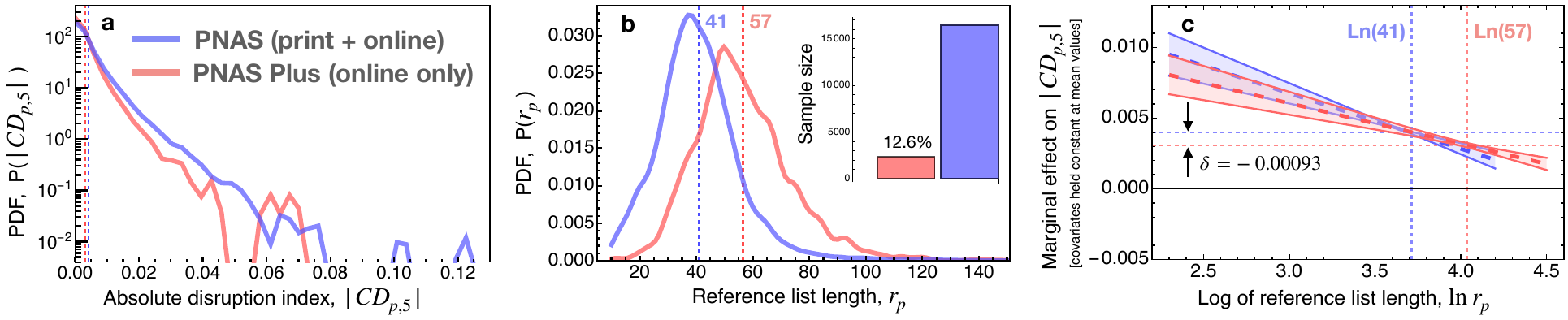}}
 \caption{ \label{Figure3.fig}  {\bf  Quasi-experimental test and validation of the CI hypothesis: counterfactual juxtaposition of  research articles   published in  {\it PNAS}    versus  {\it PNAS Plus}.}  
 {\bf (a)} Frequency distribution of the absolute disruption index, $\vert CD_{p,5}\vert$.
 {\bf (b)} Frequency distribution of the number of references per paper, $ r_{p}$.
See {\bf Fig. \ref{FigureS3.fig}} for comparison of the two subsamples across a wider range of characteristics.
 Dashed vertical bars indicate the subsample means. 
 {\bf (c)} For both subsamples, the decline in is fully attributable to the variation in $r_{p}$ such that the difference in average reference list lengths accounts for the entire, albeit small, difference in average $\vert CD_{p,5}\vert$. 
% For (c) margins extended over the  95th percentile range of rp for each subset
}
\end{figure*}

Because computing $CD_{p,5}$ requires 5 years of post-publication citation data, in what follows we compare the disruptiveness of 18,644 research articles  published in PNAS from 2011-2015. Notably, online-only {\it PNAS Plus} articles feature a different page numbering system, and so by inspecting this metadata for each article we identified  12.6\% of the total sample as {\it PNAS Plus} articles. While the difference in the average $\vert CD_{p,5}\vert$ is incremental, the {\it PNAS Plus} articles do have smaller disruption values in magnitude across the bulk of the sample distribution --  see  {\bf  Fig. \ref{Figure3.fig}}(a,b). 
In terms of relevant citation network characteristics related to $CD_{p}$,  {\it PNAS Plus} articles differ primarily in terms of  $r_{p}$, as they feature $100 \times (57 - 41)/41 = 39\%$ more references per article, on average.
Otherwise,  the two subsamples are nearly indistinguishable in terms of citation impact ($c_{p,5}$) and team size ($k_{p}$) -- see {\bf  Fig. \ref{FigureS3.fig}}. 

We exploit this quasi-experimental setting in order to distinguish between the following behavioral (i) and statistical (ii \& iii) mechanisms that could contribute to  declines in  $CD$:
\begin{enumerate}
\item[(i)] $N_{j} > N_{i}$: the main hypothesis for explaining the decline in $CD$ put forward by Park et al.  \cite{park2023papers} is that there have been  fundamental shifts in scientific practice that have shifted away from disruptive science towards consolidating science. However, they do not eliminate the possibility that $N_{i}$ is growing faster than $N_{j}$, on average,  due to behavioral shifts affecting scholarly citation practice. Hence, increasing rates of $N_{j}$ relative to $N_{i}$, may follow from a number of competing practical mechanisms, which they do discuss, but do not distinguish.
\item[(ii)] Statistical `large N' convergence of $CD$: The distribution of  $CD$ is extremely concentrated around the centroid value of 0. Hence, it is possible that $(N_{i} - N_{j}) \rightarrow 0$ as $c(t) \gg 1$ and $r(t) \gg 1$ increase over time, representing a statistical limit associated with increasing network density. Note that this candidate mechanism is reflected by the numerator of $CD$ in Eq. (\ref{eqnCD}). Such statistical convergence would explain the very small variance in the $CD_{p,5}$ distribution, which is extremely leptokurtic  -- see {\bf Fig. \ref{FigureS4.fig}}; 
\item[(iii)] Citation inflation: as $R_{k}\gg 1$, $CD$ converges to 0 (since the numerator of $CD$ is bounded). Unlike (ii), the source of this statistical mechanism is in the denominator of Eq. (\ref{eqnCD}).
\end{enumerate}
Park et al. \cite{park2023papers}  primarily  attribute the observed decline in $CD$ to shifting balance of  disruptive innovation captured by mechanism (i). 
However, they do not rule out  mechanisms (ii) or (iii), which are not related to the  innovation capacity of the scientific enterprise, but instead reflect the susceptibility of the $CD$ metric to statistical bias.
Hence, we empirically test the CI hypothesis using a simplified CD metric, $\vert CD_{p,5} \vert$, which is not sensitive to mechanism (i). This modification is not dissimilar to the choice of alternative disruption metric employed by Lin et al. \cite{Lin_remote_2023}, who base their results on the sign of $CD$, which  avoids the measurement bias associated with mechanism (iii).

We test the    relationship   between $\vert CD_{p,5} \vert$ and various covariates  using the model specification employed in our companion study \cite{petersen2023disruption} (which is based upon a  different dataset).  
Instead, here we use publicly available publication metadata from the {\it SciSciNet} open data repository \cite{lin2023sciscinet}, which features pre-calculated $CD_{p,5}$, $k_{p}$, $r_{p}$. 
We use the following  multivariate linear regression model, 
\begin{eqnarray}
\label{eqnReg}
\vert CD_{p,5} \vert  = b_{t} + b_k \ln k_{p} + b_r \ln r_{p} + b_c \ln c_{p,5}   +  \epsilon_t \ .
\end{eqnarray}
which accounts for  team size and  the most relevant scalar citation network quantities relating to $CD$.
We estimate the parameters of the model using the STATA 13 package ``xtreg fe'' using publication-year fixed effects; each covariate enters in logarithm to temper the right-skew in the distribution of each variable.
For the full list of parameter estimates see  {\bf Table \ref{PNASPlusFE.reg}}. We also tested the robustness of the parameter estimates by applying the same model to a larger  sample of journals,  comprised of 6.9 million articles from the same period, 2011-2015, which shows consistent results across a larger range of journals -- see  {\bf Table \ref{AllFE.reg}}.

Results show a negative relationship between $\vert CD_{p,5} \vert$ and $r_{p}$: $b_{r} =  -0.0039$; $p = 0.002$; 95\% CI = [ -0.0055   -0.0023] which further supports the CI hypothesis. 
Put in real terms, a paper with twice as many references ($2r_{p}$) has a $\vert CD_{p,5} \vert$ value that is $b_r \ln(2) = -0.002$ smaller than if it had $r_{p}$ references. 
This scenario corresponds to a $0.6\sigma$ effect size, as the joint standard deviation across both PNAS subsamples is $\sigma[\vert CD_{p,5} \vert]  = 0.0065$.
Notably, the sign, magnitude, statistical significance level  of $b_r$ is consistent with the analog coefficient reported in \cite{leahey2023types}. The relationship between $\vert CD_{p,5} \vert$ and $k_{p}$ are not robust in sign, which is likely attributable to the small effect size compounded by the non-linear increasing relationship between $k_{p}$ and  $r_{p}$ over time \cite{petersen2023disruption}, which we address in the following section.   

Moreover, this model facilitates estimating the differences in $\vert CD_{p,5} \vert$ between the {\it PNAS} and {\it PNAS Plus}  deriving solely from the  differences in  $r_{p}$. Our  results  show that  100\% of the difference in the average $\vert CD_{p,5} \vert $ between the two journal subsets are explained by $\delta$, the  difference  in the average $r_{p}$ across the two  subsets -- see {\bf  Fig. \ref{Figure3.fig}}(c). Hence, these empirical results  definitively demonstrate  that a significant portion of variation in $CD$ is attributable to variation in $r_{p}$. For this reason, the main results  reported in \cite{park2023papers} survived their robustness checks, e.g. the random rewiring they employed conserves $r_{p}$, and   Extended Table 1 and Supplementary Table 3 show that they did not include   $r_{p}$, $k_{p}$ or $k_{p}$ as publication-level covariates of $CD$.\\

%\bigskip
%\centerline{[Figure 2 Here]}
%\bigskip
 
\noindent {\bf Empirical analysis of  $CD$ from 1995-2015}\\
\noindent 
There is considerable disagreement emerging from research analyzing the relationships between $CD$ and various  other factors. For example,   \cite{wu2019large} mainly rely on  descriptive   methods to establish a negative relationship between  $CD_{p}$ and the team size, $k_{p}$. Instead, \cite{petersen2023disruption} and \cite{leahey2023types} employ multivariate regression and report a positive relationship, and no relationship between $CD_{p}$ and $k_{p}$, respectively. One reason for the discrepancy emerging in the literature is a lack of consistency in the  data and methodological specifications. 

Hence, in this section we re-analyze publication-level temporal trends \cite{park2023papers} and  team-size trends  \cite{wu2019large} in $CD_{p,5}$ using pre-generated and publicly available citation network data from   {\it SciSciNet}  \cite{lin2023sciscinet}. For consistency, we apply the same general model specification developed in the previous sections and also applied in \cite{petersen2023disruption}.
Furthermore, we  restrict our analysis to publications that feature explicit signatures of research outcomes  -- namely, those with sufficiently large $r_{p}$  that we can be confident that they are not editorials, commentaries, book reviews or other non-research based content that may be misclassified as such. 
This selection also excludes publications featuring substantial missing network data  \cite{macher2023illusive}, since these data quality issues effectively reduce $r_{p}$,  and consequently give rise to  spuriously large $\pm CD$; this selection also avoids the issue deriving from the  surprisingly frequent singularity identified by Holst \cite{holst2024dataset}  whereby papers with $r_{p}=0$ generate $CD_{p}=1$. 
As such, we focus on the components of the citation network that are both conceivably and consequentially   disruptive, in line with the originator's definition of disruption representing a form of breakthrough innovation  \cite{christensen2013disruptive}.
 
With this in mind, we ranked journals over the period  1995-2015 according to the number of publications with $10 \leq r_{p} \leq 200$.
We then analyze publications from the   top 1000 most prominent journals, which also satisfy the following criteria:   team sizes in the range $1 \leq k_{p} \leq 25$, and citation counts  in the range $1 \leq c_{p,5} \leq 1000$. This exclusion produces a 0.26\% decrease in the sample size, resulting in 7,819,889 publications. 
By focusing on  prominent journals, we can also calculate a journal-year normalized disruption index, 
\begin{eqnarray}
\label{NormCD}
\text{Norm}CD_{p,5,j,t} = \frac{CD_{p,5,j,t}  - \overline{CD}_{j,t}}{\sigma[CD]_{j,t}} \ ,
\end{eqnarray}
where $\overline{CD}_{j,t}$ is the average $CD$ value  and $\sigma[CD]_{j,t}$ is the standard deviation calculated for publications from journal $j$ in year $t$.
As such, this normalized disruption metric controls for year-specific  factors such as journal publication modality (online, print, hybrid), as well as the characteristic value and  variation CD according to the discipline associated with $j$, etc.
As a final data quality assurance, we exclude publications with $\vert \text{Norm}CD_{p,5,j,t} \vert \geq 5$, which corresponds to just a 0.66\% decrease in the sample size.
The resulting sample is comprised of 7,768,207 publications (comprising 99\% of the original data sample), with the most productive (least productive) journal featuring 138,883 (respectively, 3371) publications over the 21-year period.   

Following these  data quality refinements, we then re-analyze the temporal trend in $CD$ using the following model specification,
\begin{eqnarray}
\label{eqnRegCDyear}
CD_{p,5,j,t}   = b_{j} + b_k \ln k_{p} + b_{k2} (\ln k_{p})^{2}  + b_{k\times t} (\ln k_{p} \times t)  +  b_r \ln r_{p} + b_{r2} (\ln r_{p})^{2} + b_c \ln c_{p,5}   + b_{c2} (\ln c_{p,5})^{2} + \gamma_{t} +  \epsilon_j \ ,
\end{eqnarray}
which incorporates squared terms to account for non-linear relationships. For example, because  $N_{k} \sim n(t) r(t)$ appears in the denominator of $CD$, a linear correction for $r_{p}$ is insufficient. The coefficient $b_{r}=-0.0033$ (p-value $<$ 0.001; 95\% CI =[-0.0042, -0.0025]) is negative, reflecting the residual impact of CI.  For the full list of parameter estimates see  {\bf Table \ref{AllFEYear.reg}}. 

\begin{figure*}
\centering{\includegraphics[width=0.99\textwidth]{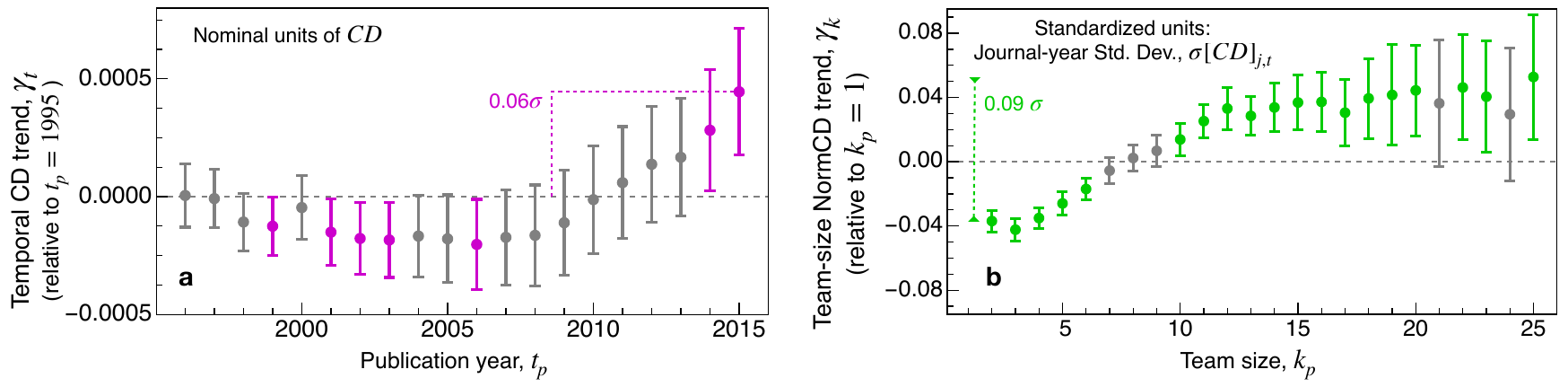}}
 \caption{ \label{Figure4.fig}  {\bf  Non-linear temporal and team-size trends in CD after controlling  for  CI.}  
 Marginal effects produced by multivariable regression that  control for  $r_{p}$ and $c_{p}$ (CI), increasing team sizes ($k_{p}$), and tendency for larger teams to produce longer papers with longer reference lists ($k_{p} \times t$).
 {\bf (a)}  Results indicate that  disruptive science has incrementally increased since 2006 -- which is consistent with three independent re-analyses reported in \cite{bentley2023disruption,macher2023illusive,holst2024dataset}. The magnitude of the  effect size ($0.06\sigma$) is relatively small.
 {\bf (b)} In contrast to \cite{wu2019large}, results indicate that large teams (incrementally) disrupt and small teams (incrementally) develop science.  
The magnitude of the   effect size ($0.09\sigma$) is inconsequential in terms of team science policy guidance and team assembly strategy.
Shown are factor variable point estimates with 95\% confidence intervals; 
Gray error bars  are not statistically deviant from the baseline level indicated by the horizontal dashed line ($p>0.05$). See {\bf Tables  \ref{AllFEYear.reg} \& \ref{AllFEteamsize.reg}} for the full list of model parameter estimates.
}
\end{figure*}

 {\bf  Figure  \ref{Figure4.fig}}(a) shows the trend in the factor variable $\gamma_{t}$, which  captures year-specific trends that persist in spite of the publication-level controls. 
 Note that the  regression adjustment robustness checks reported in Supplementary Table 1 of \cite{park2023papers} does not report any of the field-year and paper-level controls, and so it is not possible to validate our results according to their covariates; in particular, they  not include the  covariates $r_{p}$,  $k_{p}$ and $c_{p,5}$ in their model specification. 
The results of our reanalysis indicates that the residual trend in $CD(t)$ associated with time is at the level of noise, with the uptick in the regression adjusted $CD(t)$ after 2008 corresponding to just $0.06\sigma$ effect size relative to the baseline level in 1995. 

In order to evaluate team-size trends, we leverage the journal-year normalized disruption index $\text{Norm}CD$ to estimate the standardized parameters of the  model
\begin{eqnarray}
\label{eqnRegCDk}
\text{Norm}CD_{p,5,j,t}   = b_{t} + \gamma_{k} + b_r \ln r_{p} + b_{r2} (\ln r_{p})^{2} + b_c \ln c_{p,5}   + b_{c2} (\ln c_{p,5})^{2} + \epsilon_t \ .
\end{eqnarray}
As such, coefficients are measured in  units of $\sigma[CD]_{j,t}$, which facilitates assessing the relative magnitude of effect sizes. 
The interaction term represented by $(\ln k_{p} \times t )$ controls for the tendency of larger teams to produce longer papers with longer reference lists \cite{petersen2023disruption}.
After controlling for temporal variation and CI, we find that $CD$  increases (albeit weakly) with team size  (for $k_{p} \in [3,25]$) -- which is consistent with a statistically significant and positive coefficient associated with $\ln r_{p}$ identified in our companion study \cite{petersen2023disruption}. 
As with the temporal trend, the net effect is  at the level of noise,  with the difference between $k_{p} = 2$ and $k_{p} = 25$ corresponding to just a $0.09\sigma$ effect size.
These  results are in stark contrast with \cite{wu2019large}. One source of discrepancy is the methodology, as their descriptive analysis does not account for multivariable interactions. Moreover, Wu et al. base their analysis upon differentials in the {\it percentile} values of $CD_{p,5}$, which obscures the relatively small magnitude of the effect size obtained for nominal $CD$ values, which are extremely narrowly distributed around $CD=0$, as illustrated in {\bf Fig. \ref{FigureS4.fig}}.

\vspace{-0.2in}
\section*{Discussion}
\vspace{-0.2in}
A growing body of research seeks to relate $CD_{p}$ to time-dependent covariates such as team size \cite{wu2019large}, novelty \cite{leahey2023types},  the geographic dispersion of team members \cite{Lin_remote_2023}, and citation impact \cite{wang2023evaluating} -- all are quantities that have systematically increased over time.
A common pattern among these studies is a result of the form: as $X$ increases, $CD$ decreases. 
However, this class of results naturally follows from the susceptibility of correlations between $X(t)$ and $CD(t)$ to (a) temporal biases associated with the secular growth of the scientific enterprise, and (b) temporal biases associated with increasing data quality of the citation network data over time. 

Data quality issues deriving from missing citations and references is a fundamental source of error   identified by  \cite{macher2023illusive,holst2024dataset} that explains the anomalous decline in $CD$ reported by \cite{park2023papers}. 
In the re-analysis by Macher et al. \cite{macher2023illusive},  missing references at the beginning of the patent data  artificially reduce $r_{p}$ for early patents; upon correcting for their omission, which effectively increases $r_{p}$ for those early patents,   the negative trend in $CD$ largely disappears.
 Similarly, Holst et al. \cite{holst2024dataset} show that a significant source of systematic error follows from including items with $r_{p}=0$ that generate   $CD_{p}=1$ outliers. They show that these anomalies tend to occur earlier in the publication and patent datasets. Upon correcting these issues, they also show that the negative trend in $CD$ largely disappears. 
This second re-analysis also provides the full set of coefficients estimated in their regression adjustment analysis, which shows a negative correlation between $CD_{p}$ and $r_{p}$ ($\beta_{1}$ in Table S1 of \cite{holst2024dataset}). Indeed, these data quality issues give rise to the same net effect as CI. 
Beyond data quality issues, another issue are the  small effect sizes between $CD$ and covariates. 
Our re-analysis of temporal trends and team-size trends  generate effect sizes at the 0.06$\sigma$ and $0.09\sigma$ level, respectively; moreover, the directions of the trends are  opposite of what was previously reported  \cite{park2023papers,wu2019large}.

To summarize, even in the absence of data quality issues, the $CD$ index  decreases over time due to two mechanisms unrelated to innovation -- one behavioral, and the other structural. 
Importantly, the disruption index  does not account for confounding shifts in   citation behavior (e.g. self-citation, impact factor boosting) that increase the rate of triadic closure  measured by $N_{j}$ in the numerator of $CD$. Thus, decreases in $CD$ could follow from a number of competing mechanisms, some behavioral, others statistical in nature.
 Hence, in this work we  began by   testing the `CI hypothesis' that underlies the structural mechanism. Indeed, shifts in  strategic behavior and normative practice are challenging to directly measure. For this reason, we  confront this issue via computational simulation in our companion work \cite{petersen2023disruption}. And in order to guide the development of unbiased citation-network metrics, we  make available an ensemble of synthetic citation networks   so  that they can be used to test future  citation-based indices for systematic bias \cite{DryadDisruption2023}.

  In short, our mixed method approaches consistently demonstrate  that CI causes  the denominator of  $CD$ defined in Eq. (\ref{eqnCD}) to systematically increase as reference lengths increase over time, which causes $CD$ to converge to 0.
 According to its present definition, there is no clear way to correct for this dependence, since $CD_{p}$ is non-linearly   related to  $r_{p}$ via the factor $N_{k}$. 
 This susceptibility is illustrated in {\bf Fig. \ref{Figure1.fig}}, which shows how a publication (or patent) needs to only cite one highly-cited publication for $N_{k}$ to increase to the extent that $CD\rightarrow 0$ independent of the difference $N_{i}-N_{j}$. The likelihood and magnitude of this one-off mechanism is  increasing over time as a result of CI \cite{pan2016memory}.
 Yet this issue even affects publications from the same cohort that have significantly different $r_{p}$. 
As a case example, we juxtaposed the disruptiveness  of PNAS versus PNAS Plus articles published from 2011-2015, which differ primarily in their article lengths. 
Results show that nearly all of the difference in disruptiveness is attributable to the PNAS Plus articles having larger $r_{p}$ on account of their extended online-only publication format.
Hence, a significant amount of the variation in $CD$ derives from variation in $r_{p}$, which could follow simply from journal-specific constraints on article lengths.
By way of example, our analysis based upon normalized shows that the covariate with the largest effect size is   $r_{p}$, which features a $0.14\sigma$ effect size for each unit change in $\ln r_{p}$ -- see {\bf Table \ref{AllFEteamsize.reg}}. 

We conclude by suggesting a policy consideration  for managing the scientific publishing enterprise, namely caps on reference list lengths to temper the effects of CI in research evaluation (see  \cite{pan2016memory} for computational modeling that instead explores the implications of a sudden increase in $r(t)$, i.e. simulating the emergence of online-only mega-journals and their impact on the citation network). Such limits on $r_{p}$ appear to cure the systematic decrease in $CD(t)$ \cite{petersen2023disruption}, and could  simultaneously address other shortcomings in citation practice such as surgical  self-citation  by authors \cite{ioannidis2019standardized} and institutional collectives \cite{tang2015there},  and journal impact-factor boosting \cite{martin2016editors,ioannidis2019user}.

% {\small
 %\vspace{-0.2in}
%\section*{Materials and Methods}
%}
 
% \vspace{-0.2in}
%\section*{Acknowledgments}
% \vspace{-0.2in}
%\noindent  We thank 

 \vspace{-0.2in}
\section*{Author contributions}
 \vspace{-0.2in}
\noindent AMP designed the research, performed the research, participated in the writing of the manuscript, collected, analyzed, and visualized  data. 
FA and FP  contributed to  the manuscript. AMP and FP designed the research.

 \vspace{-0.2in}
\section*{Competing  interests}
 \vspace{-0.2in}
\noindent We declare no competing financial interests. 

 \vspace{-0.2in}
\section*{\bf Data and materials availability}
 \vspace{-0.2in}
\noindent  Synthetic  citation networks   and code for analyzing them are  available in the  Dryad open data repository: DOI: \href{https://datadryad.org/stash/dataset/doi:10.6071/M3G674}{10.6071/M3G674}. Data used for the  multivariate regression analysis, including the disruption index $CD_{p,5}$, $c_{p,5}$, $k_{p}$ and $r_{p}$, were obtained  from the {\it SciSciNet} open data repository \cite{lin2023sciscinet}. 

\clearpage
\newpage

\bibliographystyle{pnas}
\bibliography{Bibtex}

\begin{comment}

\end{comment}

%\end{document}

\clearpage
\newpage

\beginsupplement

\begin{widetext}

%\end{document} <<<<<<<< END MAIN DOCUMENT HERE

%\vspace{-0.2in}
\centering{\large{\bf Supplemental Information: Figures S1-S4 and Tables S1-S4}}
\vspace{-54.2in}

\begin{figure*}[b!]
\centering{\includegraphics[width=0.59\textwidth]{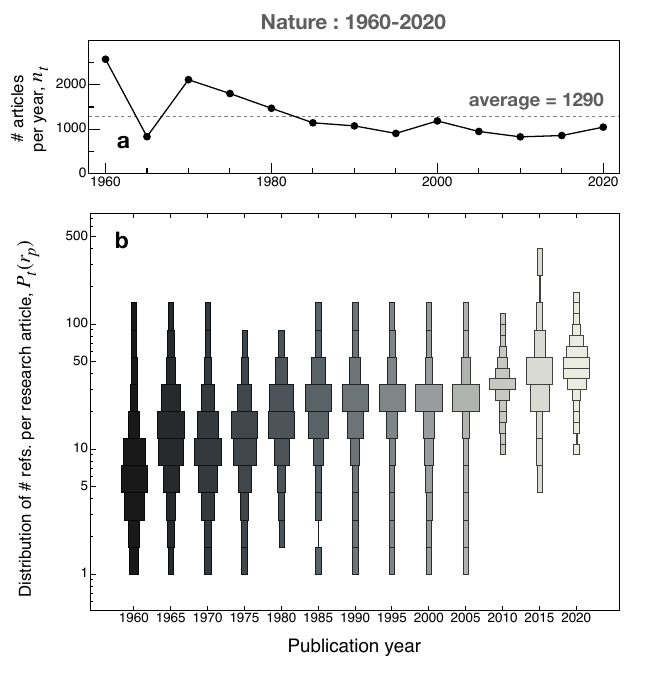}}
 \caption{  \label{FigureS1.fig} {\bf Citation inflation affects journals of all sizes, even those with relatively small change in $n(t)$.}
{\bf (a)} The number of research articles, $n_{t}$,  published by {\it Nature} tabulated over 5-year intervals from 1960 to 2020. Counts are based upon Clarivate Analytics Web of Science Core Collection, using records classified as document type =  ``Article'' and neglecting articles with $r_{p}=0$, which are likely misclassified editorial comments and the like. 
 {\bf (b)} The frequency distribution $P_{t} (r_{p})$ shows the distribution  of the number of references per article, and indicates that the increasing trend in {\bf Figure 1}(d) is not attributable to outliers, but rather a systematic shift towards larger $r_{p}$ values.  
}
\end{figure*}

\begin{figure*}
\centering{\includegraphics[width=0.72\textwidth]{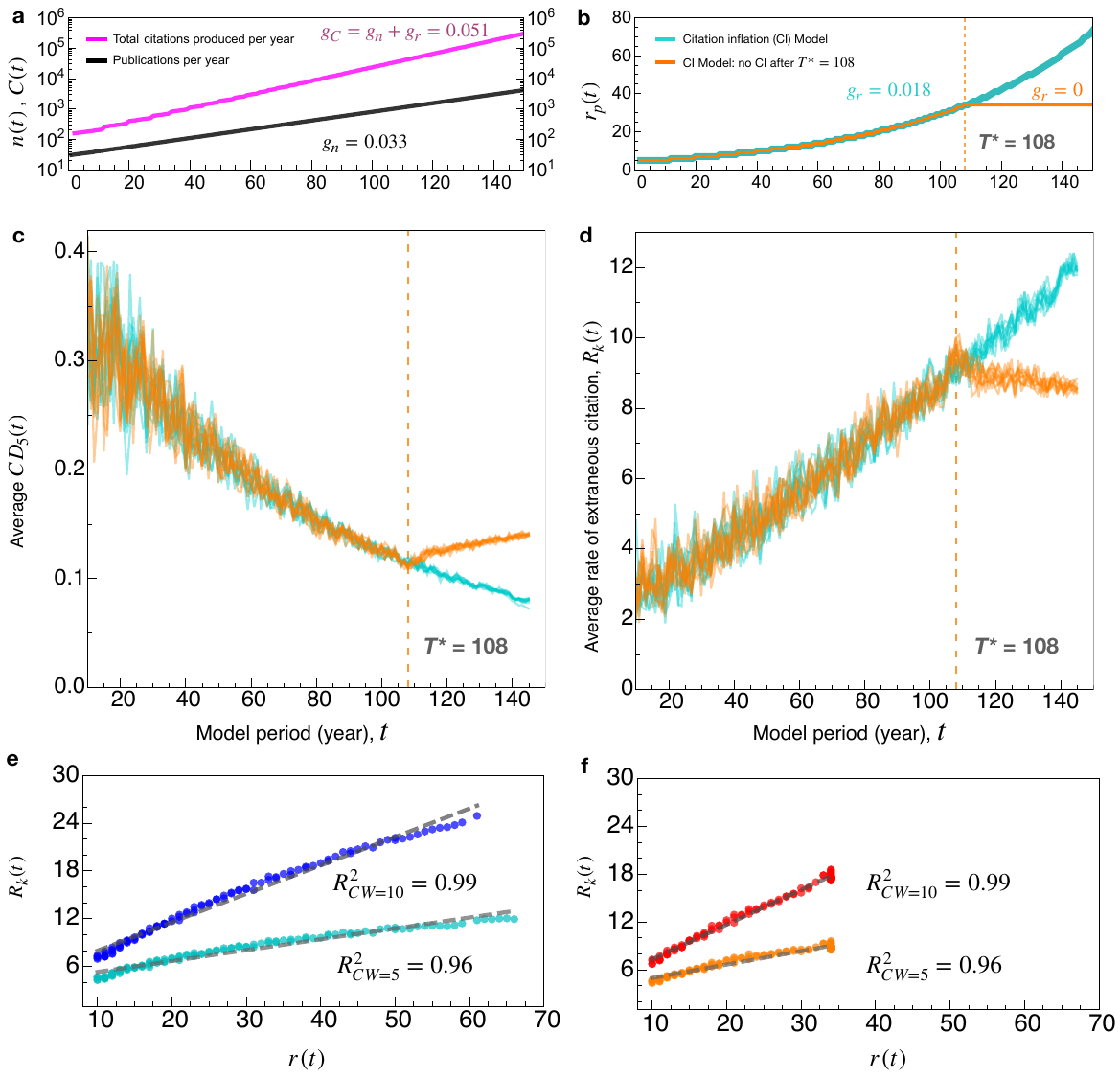}} 
\caption{ \label{FigureS2.fig} {\bf Computational simulation of growing citation networks elucidates roles of citation inflation  underlying the decline of CD.} 
 {\bf (a)} Model system evolved over $T=150$  periods (representing years). In each period, the number of new publications, $n(t)$, and the reference list length, $r(t)$, are deterministically specified, however the particular papers that each new publication cites are stochastically generated according to a network model that reproduces various structural and dynamic characteristics of real citation networks -- see \cite{pan2016memory,petersen2023disruption} for more model details. 
 Most importantly, the growth of  $n(t)$ and $r(t)$ are matched to the real growth rates, $g_{n}=0.033$ and $g_{r}=0.018$, 
 estimated using the Clarivate Analytics  Web of Science citation network  \cite{pan2016memory}. 
 As such, the total number of citations produced per year, $C(t)=n(t)r(t)$, grows at compounded exponential rate of roughly 5\% per year  ($g_{C} = g_{n}+g_{r} = 0.051$), which 
 means that the total number of links in the citation network doubles every $\ln(2)/g_{C} = 13.6$ years (or periods in our model).  
  {\bf (b)} In order to demonstrate how citation inflation deriving from increasing $r(t)$ affects the measurement of $CD_{5}(t)$, we constructed  synthetic networks that share 
  the same model parameters except for $g_{r}$: in scenario (a) we grow citation networks with fixed $g_{r} = 0.018$ for the entire $T=150$ periods (shown in orange); and in scenario (b) we grow citation networks that start  with  $g_{r} = 0.018$, but is abruptly changed to $g_{r} =0 $ for $t\geq T^{*}=108$, which halts the growth  of reference lists, such that $r(t \vert t \geq T^{*})= r(t=108) = 34$ (cyan). Scenario (b) thus explores a hypothetical publishing policy whereby journals place a  cap on reference list lengths. For example, {\it Nature} presently provides a soft policy cap  -- \href{https://www.nature.com/nature/for-authors/formatting-guide}{``As a guideline, articles typically have no more than 50 references'' (link)}.
    {\bf (c)} Each  curve shows the average $CD_{5}(t)$  calculated for a distinct synthetic network. 
    To demonstrate robustness across an ensemble of  synthetic networks generated with uniform parameters, we show the average $CD_{5}(t)$ for each of 10 different synthetic network realizations per scenario. Notably, each synthetic citation network shares  common initial conditions, independent of the scenario, and also share the same growth parameters for $t\leq 108$.
    Hence, the differences in average $CD_{5}(t)$  for $t<T^{*}$ are  attributable only to stochastic dynamics, and this variation is relatively small  across the 20 network realizations. 
   Conversely, for $t\geq T^{*}$ the differences between the scenarios is primarily attributable to citation inflation deriving from $g_{r}>0$ relative to the counterfactual scenario, $g_{r}=0$.  Interestingly, for all  realizations featuring $g_{r}=0$, the average $CD_{5}(t)$ increases over time.
    {\bf (d)} $R_{k}(t)$ is the average rate of extraneous citations, which appears to be an extensive quantity, growing proportional to the system size in scenario (a). However, in scenario (b),  $R_{k}(t)$ abruptly reverses course, and appears to saturate in the long term. 
 {\bf (e,f)}  The proportionality between $R_{k}(t)$ and $r(t)$ is one of the fundamental assumptions of our critique, as it connects citation inflation deriving from $r(t)$ to the systematic convergence of $CD_{p} \rightarrow 0$. We validate this assumption using this citation network growth model; we also validate this relationship empirically using a comprehensive dataset comprised of 30 million research articles published over the period 1945-2012  -- see \cite{petersen2023disruption}. 
 To demonstrate the robustness of this relationship, we also calculated  $CD_{p,CW}$ using citation windows (CW) of  5 and 10 periods.
 The proportionality between  $R_{k}(t)$ and $r(t)$ persists for both $CW$ used, which demonstrates  that fixed citation windows  do not address the fundamental secular growth bias associated with citation inflation.
}
\end{figure*}

\begin{figure*}
\centering{\includegraphics[width=0.79\textwidth]{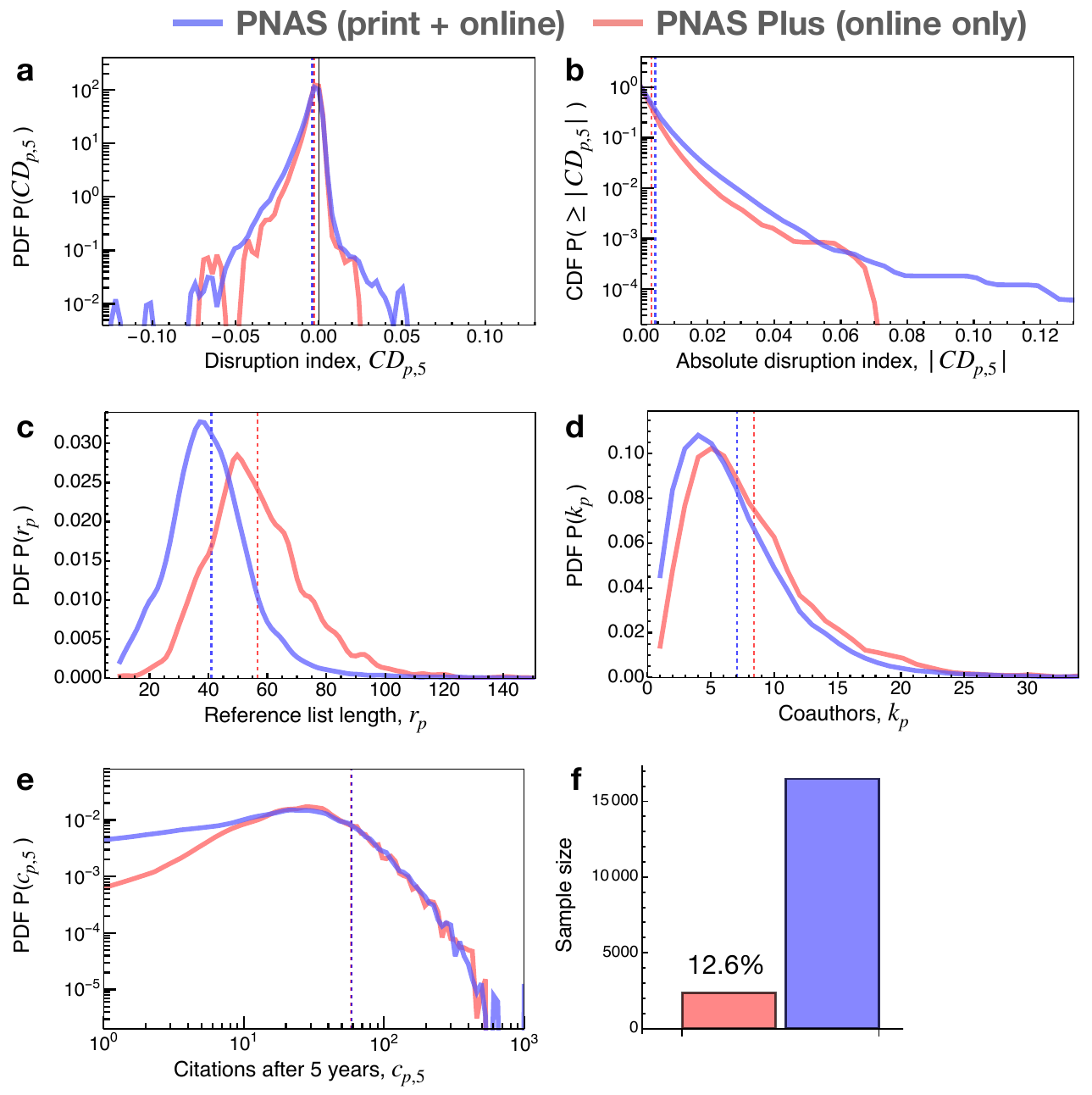}}
 \caption{  \label{FigureS3.fig} {\bf Similarity of standard (print and online) {\it PNAS}  and online-only {\it PNAS Plus} research articles.} 
 Each research article sample includes all publications from the inaugural year of the online-only {\it PNAS Plus} option in 2011 \cite{schekman2010creating} through 2015, since publications from this final cohort need  5 years of post-publication data to calculate  $CD_{p,5}$.
  {\bf (a-e)} Dashed vertical bars indicate the subsample means. 
 {\bf (a)} Frequency distribution of $CD_{p,5}$.
 {\bf (b)} Frequency distribution of the absolute value, $\vert CD_{p,5}\vert $.
 {\bf (c)} Frequency distribution of the number of references per paper (i.e., the ref. list length), $ r_{p}$, which is the variable featuring the most significant difference between the two samples.
  {\bf (d)} Frequency distribution of the number of coauthors, $k_{p}$.
   {\bf (e)} Frequency distribution of the number of citations received within a  5-year window, $c_{p,5}$.
    {\bf (f)} Sample sizes, with {\it PNAS Plus} accounting for roughly 12.6\% of the total sample size $N=18,644$ publications.
    All quantities analyzed were obtained from the {\it SciSciNet} open data repository \cite{lin2023sciscinet}.
}
\end{figure*}

\begin{figure*}
\centering{\includegraphics[width=0.65\textwidth]{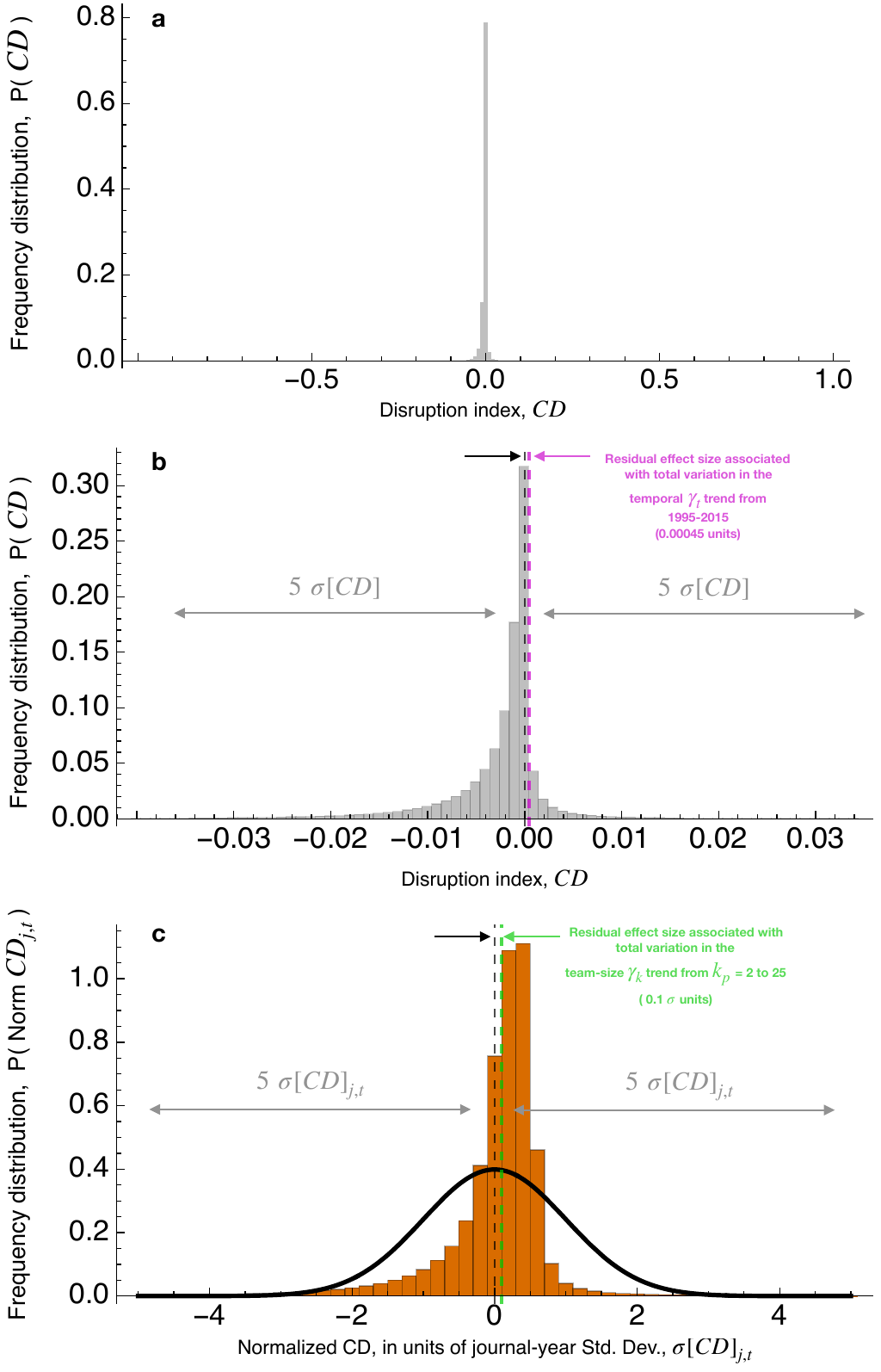}}
 \caption{  \label{FigureS4.fig} {\bf Descriptive statistics: extremely concentrated distribution of CD.}
Publication data   were obtained from the {\it SciSciNet} open data repository \cite{lin2023sciscinet}, selecting journal articles published in the 21-year range 1995-2015, and further selecting only the articles with  reference list lengths in the range $10 \leq r_{p} \leq 200$. This assures that we are not basing results upon editorials, comments, and other articles that do not represent substantial research products, and avoids the susceptibility of $CD$ to small data sample fluctuations, e.g. $CD=  N_{i} / N_{i}  = 1$ if $r_{p} = 0$ and $c_{p}>0$.  
We further select articles with $1 \leq c_{p,5} \leq 1000$ and $1 \leq k_{p} \leq 25$, which reduce the  sample size by just 0.26\%.
And finally, we  select articles with $\vert \text{Norm}CD_{p,5} \vert \leq 5$, which reduces the sample by 0.66\%. Hence, in combination the $k_{p}$, $c_{p}$ and $\text{Norm}CD_{p,5}$ thresholds reduce the sample size by only 1\%, and so the resulting analyses are based upon 99\% of the data, which captures trends based upon the overwhelmingly vast majority of research articles.
{\bf (a)} The frequency distribution  $P(CD_{p,5})$ is extremely leptokurtic.
 {\bf (b)} Zooming in on the $\pm 5\sigma$ range of the $CD_{p,5})$ distribution; 95.5\% of the data are within $\pm 2\sigma$ of the average CD value. % (-0.00266454). 
 The deviation between the vertical black and magenta lines shows the relatively small amount of variation in CD attributable to the temporal trends shown in {\bf Fig. \ref{Figure4.fig}}(a), which shows $\gamma_{t}$ from 1995-2015 (a total of 0.00045 units of CD, corresponding to $0.06\sigma$ in terms of the standard deviation of  $CD$). % 
 {\bf (c)} Frequency distribution  $P(\text{Norm}CD_{p,5})$ over the $\pm 5\sigma$ range. The deviation between the vertical black and green lines shows the $0.09 \sigma$ effect size attributable to the team-size trends shown in {\bf Fig. \ref{Figure4.fig}}(b). For visual comparison, the solid black curve represents a $N(0,1)$ normal distribution, showing that $CD$ is still leptokurtic after accounting for journal-year specific location and scale.
}
\end{figure*}

\clearpage
\newpage

\begin{table*}
\caption{\label{PNASPlusFE.reg} {\bf Quasi-experimental test of the CI hypothesis.} 
Results of multivariate linear regression juxtaposing research published in {\it PNAS} (print and online) versus {\it PNAS Plus} (online-only). Regression  implemented using the same model specification indicated by  Eqn. (\ref{eqnReg}). 
Below each coefficient estimate is   the   standard error  shown in parentheses. The first four columns show partial models, and the fifth shows the full  multivariate model. The sixth column shows the results of the full model with $D_{PNAS+}$ substituting for $\ln r_{p}$, such that the coefficient explaining the differences in $\vert CD_{p,5} \vert$ attributable to the difference in means between the two samples is $\delta =   -0.000947$, which is precisely the difference indicated by  the interaction model (6) and the corresponding marginal effects shown  in {\bf Fig. 2}.} %kmax = 25
\resizebox{0.85\columnwidth}{!}{ % ADD WRAPPER  << REPLACE FIRST {
\def\sym#1{\ifmmode^{#1}\else\(^{#1}\)\fi}
\begin{tabular}{l*{7}{c}}
\hline\hline
  Dependent variable            &\multicolumn{5}{c}{$\vert CD_{p,5} \vert$}\\
   %Dependent variable            &\multicolumn{1}{c}{$\vert CD_{p,5} \vert$}&\multicolumn{1}{c}{$\vert CD_{p,5} \vert$}&\multicolumn{1}{c}{$\vert CD_{p,5} \vert$}&\multicolumn{1}{c}{$\vert CD_{p,5} \vert$}&\multicolumn{1}{c}{$\vert CD_{p,5} \vert$}&\multicolumn{1}{c}{$\vert CD_{p,5} \vert$}\\
     &\multicolumn{1}{c}{(1)}&\multicolumn{1}{c}{(2)}&\multicolumn{1}{c}{(3)}&\multicolumn{1}{c}{(4)}&\multicolumn{1}{c}{(5)}&\multicolumn{1}{c}{(6)}&\multicolumn{1}{c}{(Full model)}\\
\hline
{\it PNAS Plus} (indicator), $D_{PNAS+}$      &   -0.000790\sym{**} &                     &                     &                     &                     & {\color{red} $\delta =$  -0.000947\sym{***}}  &     {\color{red}    -0.00154  }     \\
            &  (0.000117)         &                     &                     &                     &                     & (0.0000682)      &        (0.00129)        \\
[1em]
Reference list length, $\ln r_{p}$     &                     &    -0.00159\sym{**} &                     &                     & {\color{magenta}   -0.00374\sym{**} } &          &       -0.00392\sym{**}                 \\
            &                     &  (0.000344)         &                     &                     &  (0.000526)         &  &          (0.000567)              \\
[1em]
\ \ \ \ \ \ \ \ \ \  $D_{PNAS+} \times  \ln r_{p}$ &              &              &               &         &             &        &      0.00106         \\
&              &              &               &         &             &        &      (0.000499)         \\
[1em]
Team size, $\ln k_{p}$       &                     &                     &    0.000807\sym{**} &                     & {\color{magenta}   -0.0000615}         &   -0.000303\sym{**}    &  -0.0000344          \\
            &                     &                     & (0.0000975)         &                     & (0.0000380)         & (0.0000573)      & (0.0000512)              \\
[1em]
\ \ \ \ \ \ \ \ \ \ $D_{PNAS+} \times  \ln k_{p}$ &              &              &               &         &             &        &   -0.000350         \\   
&              &              &               &         &             &        &    (0.000203)          \\       
[1em]
Citation impact, $\ln c_{p,5}$     &                     &                     &                     &     0.00234\sym{***}&   {\color{magenta}   0.00284\sym{***}}&     0.00243\sym{***}  &     0.00290\sym{***}         \\
            &                     &                     &                     &  (0.000213)         &  (0.000268)         &  (0.000221)     &       (0.000267)        \\
[1em]
\ \ \ \ \ \ \ \ \ \ $D_{PNAS+} \times  \ln c_{p,5}$ &              &              &               &         &             &        &   -0.000474         \\
&              &              &               &         &             &        &   (0.000228)         \\
[1em]
Constant      &     0.00419\sym{***}&     0.00996\sym{**} &     0.00270\sym{***}&    -0.00449\sym{**} &     0.00755\sym{**} &    -0.00417\sym{**}   &   0.00795\sym{**}       \\
            & (0.0000146)         &   (0.00127)         &  (0.000169)         &  (0.000782)         &   (0.00101)         &  (0.000749)      &       (0.000228)           \\
\hline
 Year   FE    &        Y       &     Y          &        Y       &     Y     &        Y       &     Y     &     Y       \\
\hline
\(N\)       &       18,644         &       18,644         &       18,644         &       18,644         &       18,644         &       18,644        &   18,644          \\
adj. \(R^{2}\)&       0.002         &       0.009         &       0.008         &       0.118         &       0.165         &       0.122       &     0.165          \\
%F           &       45.89         &       21.32         &       68.49         &       120.4         &       464.3         &       84.31  &  .         \\
%df\_m        &           0         &           0         &           0         &           0         &           2         &           2   & 3       \\
%df\_r        &           4         &           4         &           4         &           4         &           4         &           4        & 4 \\
\hline\hline
\multicolumn{8}{l}{\footnotesize \textit{p}-values in parentheses}\\
\multicolumn{8}{l}{\footnotesize \sym{*} \(p<0.05\), \sym{**} \(p<0.01\), \sym{***} \(p<0.001\)}\\
\end{tabular}
}
\end{table*}

\begin{table*}\centering
\caption{\label{AllFE.reg} {\bf Robustness check using  a comprehensive data sample comprised of 6.9 million journal articles from 2011-2015.} 
Results of multivariate linear regression  implemented in STATA 13 for dependent variable $\vert CD_{p,5} \vert$,  controlling for $r_{p}$ and secular growth by way of yearly fixed-effects, as specified in Eqn. (\ref{eqnReg}). 
Publication data   were obtained from the {\it SciSciNet} open data repository \cite{lin2023sciscinet}, selecting journal articles published in the 5-year range 2011-2015, with reference list lengths in the range $10 \leq r_{p} \leq 200$, with team sizes in the range $1 \leq k_{p} \leq 25$, and citation count within 5 years of publication in the range $1 \leq c_{p,5} \leq 1000$. 
 Covariates are included following a logarithmic transform. Shown below each coefficient estimate is   the   standard error  (in parentheses). The first three columns show partial models, and the fourth shows the full  multivariate model; column model numbers coincide with those in {\bf Table \ref{PNASPlusFE.reg}}. %kmax = 25
}
\resizebox{0.6\columnwidth}{!}{ % ADD WRAPPER  << REPLACE FIRST {
\def\sym#1{\ifmmode^{#1}\else\(^{#1}\)\fi}
\begin{tabular}{l*{6}{c}}
\hline\hline
  Dependent variable            &\multicolumn{4}{c}{$\vert CD_{p,5} \vert$}\\
 &\multicolumn{1}{c}{(2)}&\multicolumn{1}{c}{(3)}&\multicolumn{1}{c}{(4)}&\multicolumn{1}{c}{(5)}\\

\hline
Reference list length, $\ln r_{p}$     &    -0.00149\sym{***}&                     &                     &   {\color{magenta}  -0.00287\sym{***}}\\
            & (0.0000557)         &                     &                     &  (0.000120)         \\
[1em]
Team size, $\ln k_{p}$       &                     &    0.000256\sym{**} &                     & {\color{magenta}   -0.000254\sym{***}}\\
            &                     & (0.0000373)         &                     &(0.00000420)         \\
[1em]
Citation impact, $\ln c_{p,5}$     &                     &                     &     0.00116\sym{***}&   {\color{magenta}   0.00176\sym{***}}\\
            &                     &                     & (0.0000731)         & (0.0000944)         \\
[1em]
Constant      &     0.00756\sym{***}&     0.00209\sym{***}&   -0.000138         &     0.00874\sym{***}\\
            &  (0.000191)         & (0.0000501)         &  (0.000162)         &  (0.000206)         \\
\hline
 Year   FE    &        Y       &     Y          &        Y       &     Y           \\
\hline
\(N\)       &     6,897,363         &     6,897,363         &     6,897,363         &     6,897,363         \\
adj. \(R^{2}\)&       0.021         &       0.001         &       0.052         &       0.119         \\
%F           &       718.5         &       46.99         &       251.8         &      3176.2         \\
%df\_m        &           0         &           0         &           0         &           2         \\
%df\_r        &           4         &           4         &           4         &           4         \\
\hline\hline
\multicolumn{5}{l}{\footnotesize Standard errors in parentheses}\\
\multicolumn{5}{l}{\footnotesize \sym{*} \(p<0.05\), \sym{**} \(p<0.01\), \sym{***} \(p<0.001\)}\\
\end{tabular}
}
\end{table*}

\begin{table*}\centering
\caption{\label{AllFEYear.reg} {\bf Non-linear cross-temporal trends in $CD$ shows increasing disruptiveness since 2011 relative to 1995.} 
Data sample are 7.8 million  articles from the top 1000 most productive journals over the period 1995-2015.
Results of multivariate linear regression  implemented in STATA 13 for  $ CD_{p,5}$ based upon the model specification in Eqn. (\ref{eqnRegCDyear}), which incorporates journal-level fixed-effects to control for unobserved time-independent factors, such as disciplinary orientation. 
Publication data   were obtained from the {\it SciSciNet} open data repository \cite{lin2023sciscinet}, and are restricted to articles with reference list lengths in the range $10 \leq r_{p} \leq 200$, with team sizes in the range $1 \leq k_{p} \leq 25$, and citation count within 5 years of publication in the range $1 \leq c_{p,5} \leq 1000$. 
 Shown below each coefficient estimate is   the   standard error  (in parentheses). 
 %The first three columns show partial models, and the fourth shows the full  multivariate model. 
 The baseline group for the year factor variable $\gamma_{t}$ is 1995. %kmax = 25
}
\resizebox{0.5\columnwidth}{!}{ % ADD WRAPPER  << REPLACE FIRST {
\def\sym#1{\ifmmode^{#1}\else\(^{#1}\)\fi}
\begin{tabular}{l*{6}{c}}
\hline\hline
  Dependent variable            &\multicolumn{4}{c}{$CD_{p,5}$}\\
 &\multicolumn{1}{c}{(1)}&\multicolumn{1}{c}{(2)}&\multicolumn{1}{c}{(3)}&\multicolumn{1}{c}{(4)}\\

\hline
Reference list length, $\ln r_{p}$     &    -0.00331\sym{***}&                     &                     &    -0.00332\sym{***}\\
            &  (0.000495)         &                     &                     &  (0.000434)         \\
[1em]
  $(\ln r_{p})^{2}$&    0.000504\sym{***}&                     &                     &    0.000654\sym{***}\\
            & (0.0000663)         &                     &                     & (0.0000576)         \\
            [1em]
Citation impact, $\ln c_{p,5}$    &                     &   -0.000154\sym{**} &                     &   -0.000154\sym{*}  \\
            &                     & (0.0000570)         &                     & (0.0000611)         \\
[1em]
$(\ln c_{p,5})^{2}$&                     &   -0.000299\sym{***}&                     &   -0.000331\sym{***}\\
            &                     & (0.0000136)         &                     & (0.0000147)         \\
[1em]
Team size, $\ln k_{p}$        &                     &                     &     -0.0543\sym{***}&     -0.0471\sym{***}\\
            &                     &                     &   (0.00729)         &   (0.00668)         \\
[1em]
$(\ln k_{p})^{2}$ &                     &                     &   0.0000530         &    0.000151\sym{***}\\
            &                     &                     & (0.0000276)         & (0.0000222)         \\
[1em]
time-dependent team-size correction, $\ln k_{p} \times t$&                     &                     &   0.0000267\sym{***}&   0.0000232\sym{***}\\
            &                     &                     &(0.00000363)         &(0.00000333)         \\
[1em]
$\gamma_{t=1995}$   &           0         &           0         &           0         &           0         \\
  %          &         (.)         &         (.)         &         (.)         &         (.)         \\
[1em]
1996   &   0.0000197         &   0.0000284         &  0.00000701         &  0.00000495         \\
            & (0.0000655)         & (0.0000673)         & (0.0000664)         & (0.0000685)         \\
[1em]
1997   &  -0.0000515         &   0.0000525         &  -0.0000806         & -0.00000817         \\
            & (0.0000620)         & (0.0000616)         & (0.0000641)         & (0.0000635)         \\
[1em]
1998   &   -0.000187\sym{**} &  -0.0000103         &   -0.000237\sym{***}&   -0.000108         \\
            & (0.0000602)         & (0.0000596)         & (0.0000622)         & (0.0000618)         \\
[1em]
1999   &   -0.000250\sym{***}&   0.0000107         &   -0.000323\sym{***}&   -0.000125\sym{*}  \\
            & (0.0000601)         & (0.0000606)         & (0.0000634)         & (0.0000635)         \\
[1em]
2000   &   -0.000201\sym{**} &    0.000133\sym{*}  &   -0.000298\sym{***}&  -0.0000464         \\
            & (0.0000678)         & (0.0000674)         & (0.0000695)         & (0.0000689)         \\
[1em]
2001   &   -0.000319\sym{***}&   0.0000606         &   -0.000442\sym{***}&   -0.000151\sym{*}  \\
            & (0.0000681)         & (0.0000682)         & (0.0000735)         & (0.0000723)         \\
[1em]
2002   &   -0.000397\sym{***}&   0.0000756         &   -0.000544\sym{***}&   -0.000178\sym{*}  \\
            & (0.0000717)         & (0.0000731)         & (0.0000793)         & (0.0000775)         \\
[1em]
2003   &   -0.000429\sym{***}&    0.000110         &   -0.000600\sym{***}&   -0.000184\sym{*}  \\
            & (0.0000727)         & (0.0000758)         & (0.0000840)         & (0.0000811)         \\
[1em]
2004  &   -0.000422\sym{***}&    0.000176\sym{*}  &   -0.000613\sym{***}&   -0.000167         \\
            & (0.0000767)         & (0.0000824)         & (0.0000931)         & (0.0000890)         \\
[1em]
2005   &   -0.000402\sym{***}&    0.000214\sym{*}  &   -0.000623\sym{***}&   -0.000179         \\
            & (0.0000817)         & (0.0000872)         &  (0.000103)         & (0.0000950)         \\
[1em]
2006   &   -0.000416\sym{***}&    0.000246\sym{**} &   -0.000662\sym{***}&   -0.000203\sym{*}  \\
            & (0.0000807)         & (0.0000887)         &  (0.000106)         & (0.0000974)         \\
[1em]
2007   &   -0.000382\sym{***}&    0.000332\sym{***}&   -0.000653\sym{***}&   -0.000173         \\
            & (0.0000848)         & (0.0000932)         &  (0.000114)         &  (0.000103)         \\
[1em]
2008   &   -0.000349\sym{***}&    0.000408\sym{***}&   -0.000649\sym{***}&   -0.000164         \\
            & (0.0000862)         & (0.0000982)         &  (0.000124)         &  (0.000109)         \\
[1em]
2009   &   -0.000299\sym{***}&    0.000523\sym{***}&   -0.000624\sym{***}&   -0.000111         \\
            & (0.0000857)         &  (0.000102)         &  (0.000129)         &  (0.000113)         \\
[1em]
2010   &   -0.000174\sym{*}  &    0.000685\sym{***}&   -0.000524\sym{***}&  -0.0000127         \\
            & (0.0000869)         &  (0.000104)         &  (0.000135)         &  (0.000117)         \\
[1em]
2011   & -0.00000670         &    0.000826\sym{***}&   -0.000385\sym{**} &   0.0000594         \\
            & (0.0000870)         &  (0.000108)         &  (0.000142)         &  (0.000121)         \\
[1em]
2012   &    0.000135         &    0.000974\sym{***}&   -0.000270         &    0.000137         \\
            & (0.0000877)         &  (0.000110)         &  (0.000147)         &  (0.000125)         \\
[1em]
2013   &    0.000261\sym{**} &     0.00108\sym{***}&   -0.000174         &    0.000167         \\
            & (0.0000886)         &  (0.000111)         &  (0.000152)         &  (0.000127)         \\
[1em]
2014   &    0.000437\sym{***}&     0.00126\sym{***}&  -0.0000288         &    0.000282\sym{*}  \\
            & (0.0000896)         &  (0.000113)         &  (0.000158)         &  (0.000131)         \\
[1em]
2015   &    0.000631\sym{***}&     0.00149\sym{***}&    0.000136         &    0.000445\sym{**} \\
            & (0.0000906)         &  (0.000117)         &  (0.000165)         &  (0.000137)         \\
[1em]
Constant      &     0.00271\sym{**} &   -0.000608\sym{***}&    -0.00153\sym{***}&     0.00399\sym{***}\\
            &  (0.000933)         & (0.0000809)         &  (0.000136)         &  (0.000810)         \\
\hline
\(N\)       &     7,768,207         &     7,768,207         &     7,768,207         &     7,768,207         \\
adj. \(R^{2}\)&       0.003         &       0.048         &       0.003         &       0.055         \\
%F           &       29.50         &       104.5         &       30.33         &       99.34         \\
%df\_m        &          21         &          21         &          22         &          26         \\
%df\_r        &         999         &         999         &         999         &         999         \\
\hline\hline
\multicolumn{5}{l}{\footnotesize Standard errors in parentheses}\\
\multicolumn{5}{l}{\footnotesize \sym{*} \(p<0.05\), \sym{**} \(p<0.01\), \sym{***} \(p<0.001\)}\\
\end{tabular}
}
\end{table*}

\begin{table*}\centering
\caption{\label{AllFEteamsize.reg} {\bf Non-linear dependence of $CD$ on team size, shows increasing disruptiveness of  teams with $k_{p}\geq 8$ coauthors.} 
Data sample are 7.8 million  articles from the top 1000 most productive journals over the period 1995-2015.
Results of multivariate linear regression  implemented in STATA 13 for the journal-year normalized disruption index $\text{Norm}CD_{p,5}$, based upon the model specification in Eqn. (\ref{eqnRegCDk}), which incorporates year fixed-effects to control for unobserved shocks. 
Since $\text{Norm}CD_{p,5}$ is a standardized metric, coefficients correspond to effect sizes measured in units of journal-year standard deviations of $CD_{p,5,t,j}$.
Publication data   were obtained from the {\it SciSciNet} open data repository \cite{lin2023sciscinet}, and are restricted to articles with reference list lengths in the range $10 \leq r_{p} \leq 200$, with team sizes in the range $1 \leq k_{p} \leq 25$, and citation count within 5 years of publication in the range $1 \leq c_{p,5} \leq 1000$. 
 Shown below each coefficient estimate is   the   standard error  (in parentheses). 
 %The first three columns show partial models, and the fourth shows the full  multivariate model. 
 The baseline group for the team size factor variable $\gamma_{k}$ are solo-authored publications ($k_{p}=1$). %kmax = 25
}
\resizebox{0.35\columnwidth}{!}{ % ADD WRAPPER  << REPLACE FIRST {
\def\sym#1{\ifmmode^{#1}\else\(^{#1}\)\fi}
\begin{tabular}{l*{6}{c}}
\hline\hline
  Dependent variable            &\multicolumn{4}{c}{$\text{Norm}CD_{p,5}$}\\
 &\multicolumn{1}{c}{(1)}&\multicolumn{1}{c}{(2)}&\multicolumn{1}{c}{(3)}&\multicolumn{1}{c}{(4)}\\

\hline
Reference list length, $\ln r_{p}$     &      -0.198\sym{***}&                     &                     &      -0.142\sym{***}\\
            &    (0.0279)         &                     &                     &    (0.0334)         \\
[1em]
  $(\ln r_{p})^{2}$ &      0.0341\sym{***}&                     &                     &      0.0451\sym{***}\\
            &   (0.00400)         &                     &                     &   (0.00446)         \\
[1em]
Citation impact, $\ln c_{p,5}$     &                     &     0.00991         &                     &     -0.0136\sym{***}\\
            &                     &   (0.00608)         &                     &   (0.00354)         \\
[1em]
$(\ln c_{p,5})^{2}$ &                     &     -0.0337\sym{***}&                     &     -0.0354\sym{***}\\
            &                     &   (0.00198)         &                     &   (0.00170)         \\
[1em]
$\gamma_{k_{p}=1}$        &                     &                     &           0         &           0         \\
  %          &                     &                     &         (.)         &         (.)         \\
[1em]
2         &                     &                     &     -0.0653\sym{***}&     -0.0369\sym{***}\\
            &                     &                     &   (0.00260)         &   (0.00318)         \\
[1em]
3         &                     &                     &     -0.0818\sym{***}&     -0.0423\sym{***}\\
            &                     &                     &   (0.00268)         &   (0.00337)         \\
[1em]
4         &                     &                     &     -0.0868\sym{***}&     -0.0351\sym{***}\\
            &                     &                     &   (0.00239)         &   (0.00316)         \\
[1em]
5         &                     &                     &     -0.0895\sym{***}&     -0.0260\sym{***}\\
            &                     &                     &   (0.00291)         &   (0.00340)         \\
[1em]
6         &                     &                     &     -0.0929\sym{***}&     -0.0170\sym{***}\\
            &                     &                     &   (0.00301)         &   (0.00327)         \\
[1em]
7         &                     &                     &     -0.0933\sym{***}&    -0.00554         \\
            &                     &                     &   (0.00364)         &   (0.00375)         \\
[1em]
8         &                     &                     &     -0.0989\sym{***}&     0.00229         \\
            &                     &                     &   (0.00427)         &   (0.00390)         \\
[1em]
9         &                     &                     &      -0.108\sym{***}&     0.00677         \\
            &                     &                     &   (0.00535)         &   (0.00481)         \\
[1em]
10        &                     &                     &      -0.116\sym{***}&      0.0138\sym{**} \\
            &                     &                     &   (0.00587)         &   (0.00479)         \\
[1em]
11        &                     &                     &      -0.121\sym{***}&      0.0252\sym{***}\\
            &                     &                     &   (0.00531)         &   (0.00503)         \\
[1em]
12        &                     &                     &      -0.132\sym{***}&      0.0331\sym{***}\\
            &                     &                     &   (0.00814)         &   (0.00625)         \\
[1em]
13        &                     &                     &      -0.149\sym{***}&      0.0285\sym{***}\\
            &                     &                     &   (0.00762)         &   (0.00575)         \\
[1em]
14        &                     &                     &      -0.158\sym{***}&      0.0338\sym{***}\\
            &                     &                     &   (0.00810)         &   (0.00724)         \\
[1em]
15        &                     &                     &      -0.171\sym{***}&      0.0369\sym{***}\\
            &                     &                     &   (0.00914)         &   (0.00813)         \\
[1em]
16        &                     &                     &      -0.187\sym{***}&      0.0372\sym{***}\\
            &                     &                     &    (0.0100)         &   (0.00888)         \\
[1em]
17        &                     &                     &      -0.208\sym{***}&      0.0305\sym{**} \\
            &                     &                     &   (0.00998)         &    (0.0100)         \\
[1em]
18        &                     &                     &      -0.210\sym{***}&      0.0394\sym{**} \\
            &                     &                     &   (0.00969)         &    (0.0119)         \\
[1em]
19        &                     &                     &      -0.223\sym{***}&      0.0415\sym{*}  \\
            &                     &                     &    (0.0132)         &    (0.0150)         \\
[1em]
20        &                     &                     &      -0.247\sym{***}&      0.0444\sym{**} \\
            &                     &                     &    (0.0135)         &    (0.0135)         \\
[1em]
21        &                     &                     &      -0.244\sym{***}&      0.0363         \\
            &                     &                     &    (0.0167)         &    (0.0189)         \\
[1em]
22        &                     &                     &      -0.238\sym{***}&      0.0462\sym{**} \\
            &                     &                     &    (0.0151)         &    (0.0157)         \\
[1em]
23        &                     &                     &      -0.260\sym{***}&      0.0405\sym{*}  \\
            &                     &                     &    (0.0125)         &    (0.0167)         \\
[1em]
24        &                     &                     &      -0.277\sym{***}&      0.0295         \\
            &                     &                     &    (0.0153)         &    (0.0199)         \\
[1em]
25        &                     &                     &      -0.256\sym{***}&      0.0527\sym{*}  \\
            &                     &                     &    (0.0203)         &    (0.0187)         \\
[1em]
Constant      &       0.290\sym{***}&       0.248\sym{***}&       0.106\sym{***}&       0.290\sym{***}\\
            &    (0.0476)         &   (0.00256)         &   (0.00291)         &    (0.0604)         \\
\hline
\(N\)       &     7,768,207         &     7,768,207        &     7,768,207         &     7,768,207         \\
adj. \(R^{2}\)&       0.001         &       0.055         &       0.001         &       0.068         \\
%F           &       970.2         &      4176.7         &           .         &           .         \\
%df\_m        &           1         &           1         &          19         &          19         \\
%df\_r        &          20         &          20         &          20         &          20         \\
\hline\hline
\multicolumn{5}{l}{\footnotesize Standard errors in parentheses}\\
\multicolumn{5}{l}{\footnotesize \sym{*} \(p<0.05\), \sym{**} \(p<0.01\), \sym{***} \(p<0.001\)}\\
\end{tabular}
}
\end{table*}

\end{widetext}
\end{document}